\documentclass[11pt, a4paper, logo, copyright]{googledeepmind}

\pdftrailerid{redacted}

\makeatletter
\renewcommand\bibentry[1]{\nocite{#1}{\frenchspacing\@nameuse{BR@r@#1\@extra@b@citeb}}}
\makeatother

\usepackage{kantlipsum, lipsum}
\usepackage{dsfont}
\usepackage{gdm-colors}
\usepackage[utf8]{inputenc}   % 允许 UTF-8 输入
\usepackage{newunicodechar}   % 定义 Unicode 字符
\usepackage{amssymb}          % 提供部分符号支持
\usepackage{wasysym,marvosym}
\usepackage{ulem}
\usepackage{float}
\usepackage{longtable}
\usepackage{listings}
\usepackage{xcolor}
\usepackage{booktabs}
\usepackage{tikz}
\usetikzlibrary{shapes.geometric, arrows.meta, positioning}
\usepackage{amsmath}
\usepackage{algorithm}
\usepackage{algpseudocode}
\usepackage{listings}
\usepackage{enumitem} % For nested lists
\usepackage{lipsum}
\usepackage[most]{tcolorbox}
\definecolor{thinkcolor}{RGB}{227,196,144}
\definecolor{observecolor}{RGB}{153,201,227}
\definecolor{explorecolor}{RGB}{178,217,200}

\tcbset{
    common/.style={
		enhanced,
		arc=0mm,
		fonttitle=\large\bfseries,
		coltitle=black,
		attach boxed title to top left={xshift=0mm,
										yshift=-0.50mm},
		boxed title style={
			skin=enhancedfirst jigsaw,
			size=small,
			arc=5mm,
			bottom=0mm,
			left=8mm,
			right=18mm,
			top=1mm},
			boxrule=0pt,
			frame hidden},
    thinkstyle/.style={
		common,
		colbacktitle=thinkcolor,
		colframe=thinkcolor,
		colback=thinkcolor!40,
		borderline north={4pt}{0pt}{thinkcolor}},
	observestyle/.style={
		common,
		colbacktitle=observecolor,
		colframe=observecolor,
		colback=observecolor!40,
		borderline north={4pt}{0pt}{observecolor}}
}

\newtcolorbox{think}{thinkstyle,title=Prompt Template}
\newtcolorbox{observe}{observestyle,title=observe}
\newtcolorbox{custom}[2][gray]{
	common,
	title=#2,
	colbacktitle=#1,
	colframe=#1,
	colback=#1!40,
	borderline north={4pt}{0pt}{#1}}
\usepackage{array, multirow, tabularx, booktabs, makecell}

\definecolor{interest_colframe}{rgb}{0.8, 0.878, 0.871}
\definecolor{interest_colback}{rgb}{0.918, 0.953, 0.949}

\definecolor{tag_colframe}{rgb}{0.965, 0.898, 0.847}
\definecolor{tag_colback}{rgb}{0.988, 0.961, 0.941}

\definecolor{exp_colback}{rgb}{0.949, 0.965, 0.980}
\definecolor{exp_colframe}{rgb}{0.878, 0.922, 0.965}

\tcbset{
    promptbox/.code args={#1/#2}{
        \tcbset{
            enhanced,
            arc=0mm,
            colframe=#1, 
            colback=#2, 
            coltitle=black,
            fonttitle=\large\bfseries,
            attach boxed title to top left={xshift=0mm, yshift=-1.0mm},
            boxed title style={
                skin=enhancedfirst jigsaw,
                size=small,
                arc=3mm,
                bottom=0mm,
                left=8mm,
                right=8mm,
                top=1mm,
                colback=#1
            },
            boxrule=0pt,
            frame hidden,
            borderline north={4pt}{0pt}{#1},
        }
    }
}

\newcommand{\assignmentQuestionName}{Question} % The word to be used as a prefix to question numbers; example alternatives: Problem, Exercise
\usepackage{caption}
\usepackage{booktabs}
\usepackage{arydshln}
\usepackage{dashrule}

\usepackage[authoryear, sort&compress, round]{natbib}

\usepackage{bbding}
\usepackage[T1]{fontenc}    % use 8-bit T1 fonts
\usepackage{hyperref}       % hyperlinks
\usepackage{url}            % simple URL typesetting
\usepackage{booktabs}       % professional-quality tables
\usepackage{nicefrac}       % compact symbols for 1/2, etc.
\usepackage{microtype}      % microtypography
\usepackage{amsmath}
\usepackage{graphicx}
\usepackage{multicol}
\usepackage[nameinlink]{cleveref}
\usepackage{bbm}
\usepackage{multirow}
\usepackage{soul}
\usepackage{float}
\usepackage{wrapfig}
\usepackage{blindtext}
\usepackage{tablefootnote}
\usepackage{amsfonts}
\usepackage[flushleft]{threeparttable}
\usepackage{colortbl}
\usepackage{mathtools,amssymb}
\usepackage{bm}
\usepackage{makecell}
\usepackage{caption}
\usepackage{capt-of}
\usepackage{array}
\usepackage{calc}      % 用于测量高度
\usepackage{caption}   % 用于 \captionof 命令
\usepackage{subcaption}  % 在导言区添加
\usepackage[bottom]{footmisc}
\usepackage{fontawesome}
\usepackage{tabularx}        % 支持自动调整列宽的表格环境

\graphicspath{{Figures/}}

\title{MARS Technical Report}

\renewcommand{\today}{}

\author{\large MARS Team}

\begin{abstract}
Modern industrial recommender systems are organized as multi-stage cascades---retrieval, pre-ranking, ranking, and re-ranking---that progressively narrow millions of candidates to a personalized display list. While each stage deploys sophisticated prediction models, the pipeline-level decisions governing channel activation, candidate quotas, model selection, and score fusion remain engineer-tuned and shared across all users within a scenario, unable to adapt to individual intent and behavioral patterns.

This technical report presents MARS (Multi-stage Agentic Recommendation System), which extends personalization from individual-stage predictions to orchestration of the entire pipeline. A centralized near-line agent, built on a fine-tuned Qwen3-0.6B model, periodically generates user-specific orchestration configurations---a structured seven-dimensional action covering pipeline mode, retrieval-channel quotas, model selection, fusion weights, and reranker settings---from a formally constrained action space. Configurations are validated, atomically cached in iGraph, and consumed asynchronously; the online cascade never waits for LLM inference. Schema checks, incumbent fallbacks, and a master kill switch provide serving safeguards and rapid rollback.

The agent policy is trained through a three-stage post-training pipeline: Confidence-aware SFT distills schema-valid generation from a Qwen3.6-Max teacher and aligns to recommendation efficiency; Format-Anchored Preference Optimization (FAPO) learns action preferences via ActionMask with posterior weighting; Group Relative Policy Optimization (GRPO) refines the policy on-policy through Sandbox replay and a frozen learned reward model without exposing exploratory candidates to live traffic.

Two-week A/B tests across five Taobao marketing scenarios with over
ten million daily active users report positive point estimates for
each scenario's key metrics under the evaluated production settings.
The largest observed gains reach $+5.22\%$ in V2P and $+7.34\%$ in
GMV, with positive CTR point estimates on both evaluated click
surfaces. Approximately $20\%$ of requests are routed through a lightweight Minimal pipeline mode.
The evaluated deployment is associated with a $12$\,ms lower mean online serving-path
latency compared with the production baseline.
Full-stack logging and one-click rollback support safe, iterative
deployment.

\end{abstract}

\begin{document}
\maketitle

% \vspace{0em}
% \vspace{-1em}

% \begin{center}
%     \includegraphics[width=0.8\textwidth]{Figures/fig_abstract.pdf}
%     \caption{MARS architecture: near-line generation and validation, online execution of cached orchestrations with latest-valid cache, incumbent fallback, and kill switch, and feedback-driven SFT–FAPO–GRPO post-training.}
%     \label{fig:abstrace}
% \end{center}

\vspace{-0.7em}
\begin{figure}[h!]
    \centering
    \includegraphics[width=0.50\textwidth]{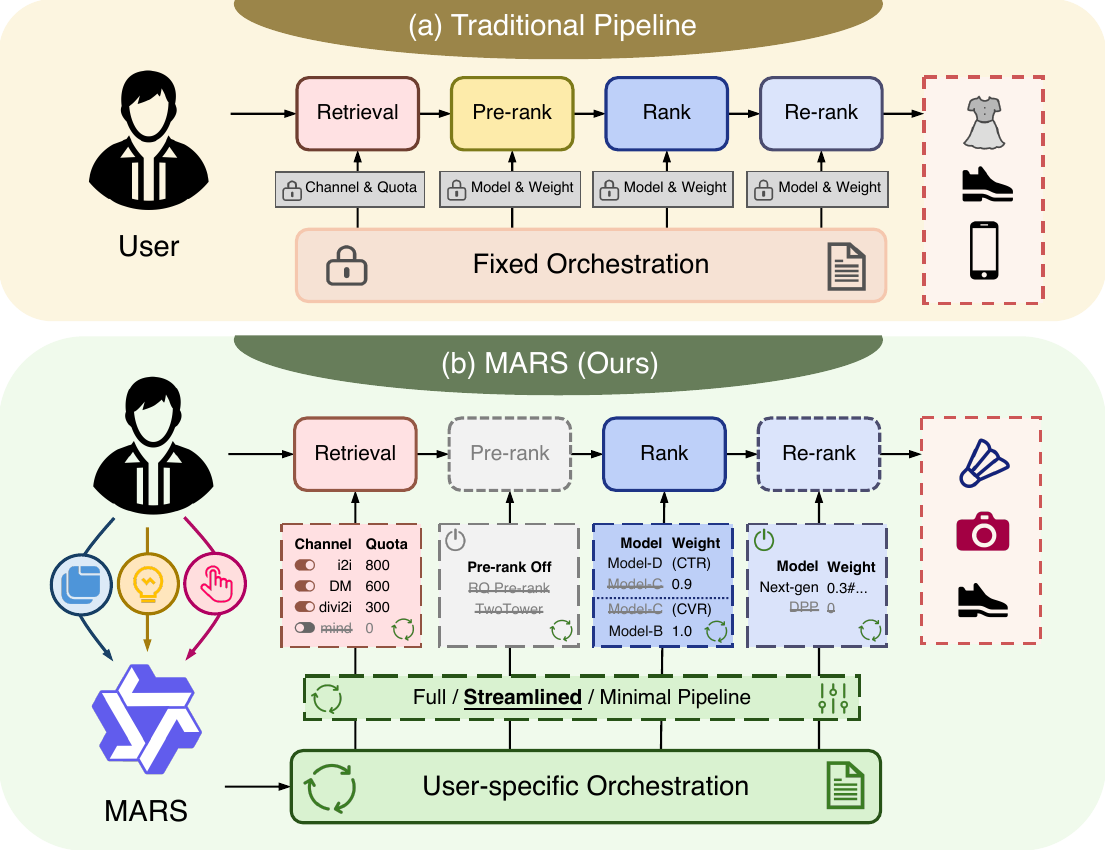}
    \vspace{-0.6em}
    \caption{Comparison of (a)~traditional fixed orchestration, where pipeline decisions are shared across all users, and (b)~MARS, which generates user-specific configurations via a near-line agent with cached serving, safety fallbacks, and three-stage post-training.}
    % \vspace{-3em}
    \label{fig:abs}
\end{figure}

\newpage
\setcounter{tocdepth}{2} % Set the depth of the table of contents to show sections and subsections only

\tableofcontents % Print the table of contents

% Incude paper content from external files
\newpage
\section{Introduction}
Large-scale industrial recommendation systems are organized as multi-stage cascades—retrieval, pre-ranking, ranking, and re-ranking—that progressively narrow millions of candidate items to a personalized display list~\citep{covington2016deep,li2019multi,zhou2018deep,guo2017deepfm,chen2018fast,pei2019personalized}. Each stage deploys increasingly sophisticated models for prediction, yet the pipeline-level decisions that govern which retrieval channels to activate, which models to invoke, how many candidates each channel contributes, and how stage-level scores are fused remain engineer-tuned and shared across all users within a scenario. This population-level static orchestration cannot adapt to the heterogeneous intent, price sensitivity, and behavioral patterns that distinguish individual users and that vary across request contexts.

Static orchestration produces three categories of mismatch. First, users in distinct interaction contexts---bargain seekers who
benefit from broad retrieval coverage, repeat purchasers who convert
best with concentrated high-relevance channels, promotion-focused
visitors with limited subsequent item engagement, and pass-through
visitors with limited interaction on the current recommendation
surface---are forced through the same pipeline configuration,
potentially leaving gains unrealized across these diverse contexts. Second, even a single user's optimal configuration is context-dependent, varying with session position, time of day, and the promotional calendar, yet static settings lack any feedback loop to adapt. Third, the combinatorial configuration space—the product of discrete channel selections, continuous weight parameters, model choices, and pipeline-mode options—is too large for manual enumeration, while the feedback for any particular configuration is delayed by hours to days, noisy, and observed only for the single configuration that was actually served, making A/B-based exploration prohibitive at the scale of millions of users and dozens of configuration dimensions.

Three recommendation-specific challenges make personalized orchestration fundamentally harder than standard configuration optimization. First, rewards are execution-dependent: a configuration cannot be evaluated without running the full pipeline, and user feedback arrives with multi-day delay, creating a credit-assignment problem that spans the entire pipeline execution and user decision process. Second, substantive decision values—channel weights, model selections, fusion parameters—occupy only a small fraction of a strictly structured response; standard preference-optimization methods treat the entire response as the unit of comparison, diluting the training signal across format tokens that should remain invariant. Third, production logs contain no counterfactual feedback for configurations that were never executed; direct online exploration is both risky and slow, so training must construct meaningful learning signals from limited observations without exposing users to unvalidated configurations.

Recent advances in instruction-tuned language models~\citep{ouyang2022training}, models with enhanced reasoning capabilities~\citep{guo2025deepseek}, and chain-of-thought prompting~\citep{wei2022chain} suggest a principled path: a model can jointly interpret user behavioral patterns, scenario constraints, and pipeline semantics, and produce an executable orchestration tailored to each request. Language models have been explored for single-stage recommendation tasks including prediction, ranking, and retrieval~\citep{geng2022recommendation,bao2023tallrec,hou2024large,lin2025can}, as well as generative retrieval and user-behavior modeling~\citep{rajput2023recommender,zhai2024actions}. However, these applications position the model within one pipeline stage rather than as an overarching orchestration policy for the entire cascade. The orchestration setting requires a valid multi-dimensional structured output, full-pipeline evaluation through actual serving infrastructure, production latency constraints, and handling of execution-dependent, sparse, and counterfactual rewards.

MARS (Multi-stage Agentic Recommendation System) addresses these challenges by strictly separating near-line generation from online execution. A centralized agent, implemented as a fine-tuned Qwen3-0.6B~\citep{yang2025qwen3}, processes user behavioral features and scenario context and generates a complete seven-dimensional orchestration configuration: pipeline mode (Full, Streamline, or Minimal), retrieval-channel subset and per-channel quotas, CTR and CVR model selections, per-stage fusion weights, and reranker settings. Generation occurs asynchronously in a near-line path; validated configurations are atomically written to a versioned iGraph cache. The online cascade reads the latest valid cache entry when serving each request and never waits for LLM inference, keeping LLM latency entirely off the critical serving path. Prior computation-allocation approaches such as
GreenFlow~\citep{lu2023greenflow}, RL-MPCA~\citep{zhou2023rl}, and
MaRCA~\citep{jiang2025marca} optimize per-stage resource budgets
within a fixed pipeline structure; MARS treats the full executable pipeline configuration as the decision, jointly selecting which stages to include, which models and channels to use, and how to combine their outputs. Among concurrent systems, DREAM~\citep{zhang2026dream} maps intent-conditioned semantic actions to guarded overrides across retrieval, ranking, and re-ranking stages; MARS generates unified configurations—including pipeline modes, channels, quotas, models, and fusion weights—as a single structured action learned through post-training.

Multiple layers of protection guard production traffic. Generation-side validation checks schema compliance, permitted model names, quota bounds, and weight constraints; only configurations that pass all checks are written to the versioned cache. Consumption-side validation independently repeats protocol and legality checks before the online cascade applies any override. Failed updates leave the latest-valid entry unchanged; when no valid cache entry exists, a read times out, or consumer-side checks fail, the system falls back to the incumbent production configuration. A master kill switch restores incumbent behavior for all traffic instantly. A diversity-exploration branch may generate and log alternative candidates for training, but these exploration candidates are never exposed to users. Full-stack logs link generated, validated, served, fallback, and explored configurations to platform-wide impressions, clicks, cart additions, purchases, and paid amounts by user ID and timestamp.

The agent policy is trained through a progressive three-stage post-training pipeline. Confidence-aware SFT distills schema-valid structured generation from a Qwen3.6-Max teacher and then aligns to recommendation efficiency using Wilson-selected execution evidence, raising format validity from 1.0\% to 99.8\%. Format-Anchored Preference Optimization (FAPO) performs offline preference learning that restricts comparison to differing action values through ActionMask, anchors full-response generation, and weights pairs by posterior visit-to-purchase rate. Group Relative Policy Optimization (GRPO) refines the policy on-policy through snapshot-based Sandbox replay and a frozen learned reward model, providing dense proxy reward signals without exposing training
candidates to users.

Two-week A/B tests across five Taobao marketing scenarios---spanning
purchase-oriented and click-oriented surfaces with more than ten
million daily active users---report positive point estimates for each
scenario's key metrics under the evaluated production settings.
The largest observed gains reach $+5.22\%$ in V2P and $+7.34\%$ in
GMV on purchase scenarios, with positive click-through-rate point
estimates on both evaluated click surfaces.

Approximately $20\%$ of requests are routed to Minimal mode based on observed
request-level patterns, including promotion-focused entry behavior
with limited subsequent item-level engagement and pass-through visits
with limited interaction with the displayed recommendation list.
This routing is associated with a reported reduction of $12$\,ms
in mean online serving-path latency under the evaluated deployment.
The system has operated continuously for three months, with
$85\%$--$95\%$ valid-cache coverage depending on the scenario.
This report makes four formal contributions. 

\faStar\;
Personalized orchestration across the recommendation pipeline: MARS extends personalization from item scoring to cross-stage orchestration—traditionally configured at the scenario level rather than per user—reuses existing capabilities, and decouples near-line generation from online execution. 

\faStar\;
Recommendation-specific post-training: confidence-aware SFT, FAPO, and GRPO form a progressive three-stage pipeline; FAPO combines a full-response generation anchor, action-level preference learning, and posterior V2P weighting, while GRPO uses snapshot-based Sandbox execution and a learned reward model. 

\faStar\;
Robust serving infrastructure: configuration validation, fallback, full-stack logging, and one-click rollback protect traffic and provide traceable feedback. 

\faStar\;
Large-scale online evaluation: two-week A/B tests across five Taobao
scenarios report positive point estimates for each scenario's key
metrics, with the largest observed gains reaching $+5.22\%$ in V2P
and $+7.34\%$ in GMV, together with a reported reduction of $12$\,ms in mean online
serving-path latency under the evaluated deployment.

%组织结构

%----------------------------------------------------------------------------------------

% \section{Overview of RecGPT} \label{sec:overview}
% \input{sections/1_5_Overview}

%----------------------------------------------------------------------------------------

\section{System Architecture}
\label{sec:overview}
MARS has three loosely coupled parts: a near-line agent that generates and validates orchestration configurations, an online cascade that executes cached orchestrations while reusing existing recommendation components through the
established configuration interface, and a logged feedback loop that supports post-training.

\begin{figure}[t]
    \centering
    \includegraphics[width=\textwidth]{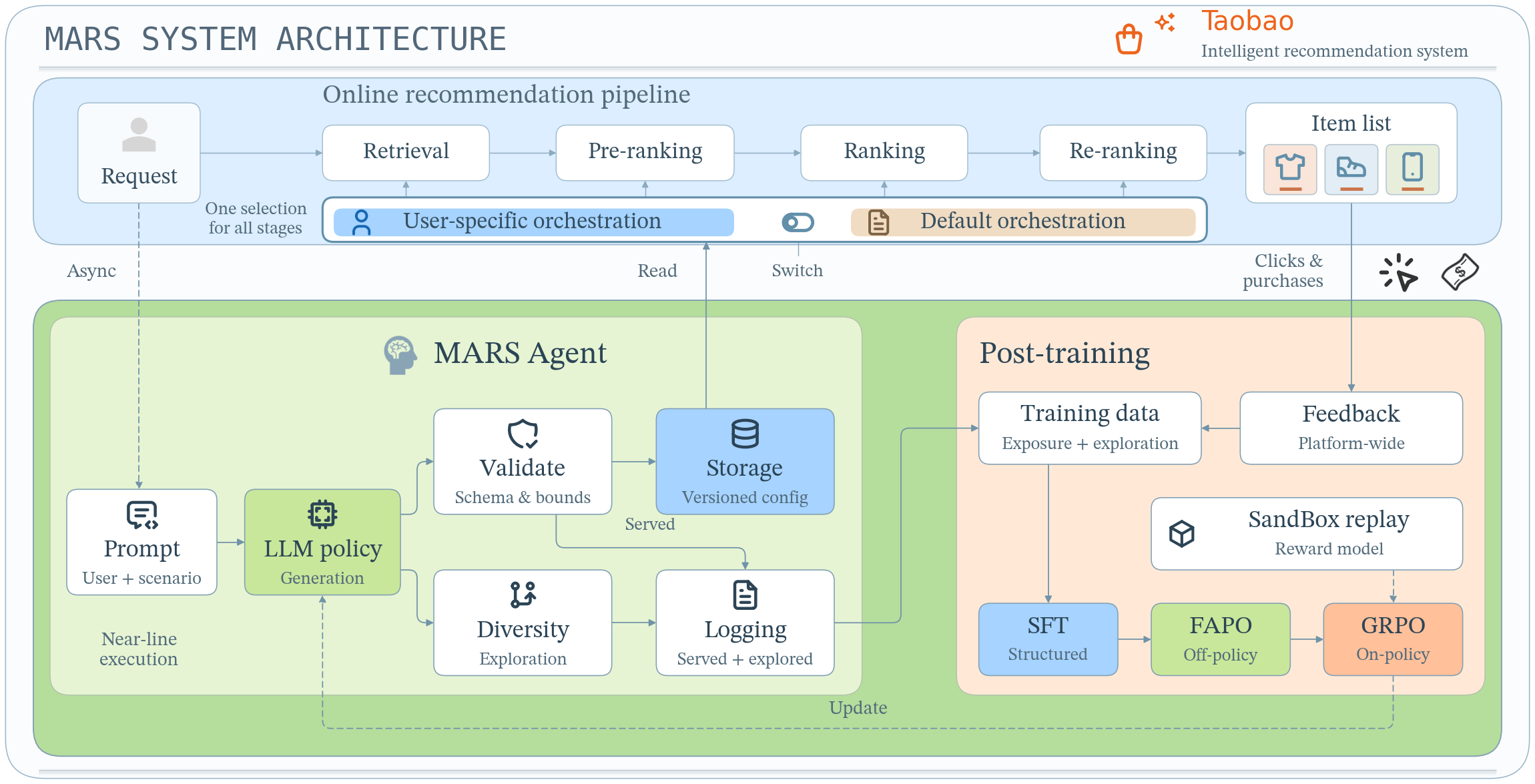}
    % \captionsetup{justification=centering}
    \caption{MARS architecture: near-line generation and validation, online execution of cached orchestrations with latest-valid cache, incumbent fallback, and kill switch, and feedback-driven SFT–FAPO–GRPO post-training.}
    \label{fig:overview}
\end{figure}

\subsection{The MARS Agent}
\label{sec:MARS_agent}
As shown in Figure ~\ref{fig:overview}, the MARS Agent operates asynchronously in a near-line path. For each request, the agent renders user behavioral features and scenario context into a structured prompt, invokes the LLM policy (a fine-tuned Qwen3-0.6B), and parses the response into a seven-dimensional orchestration configuration. Generation-side validation then checks required fields and types, permitted model names, retrieval-quota bounds, and non-negative on-grid fusion weights. Only configurations that pass all checks are atomically written to a versioned iGraph cache (Storage); a failed inference, invalid output, or timeout never overwrites the last valid version. A diversity-exploration branch may generate and log alternative candidates for post-training, but these exploration candidates are never exposed to users. The Logging module records both served and explored configurations alongside their downstream outcomes.

%----------------------------------------------------------------------------------------

\subsection{Orchestration Workflow}
\label{sec:orchestration_workflow}
For request $x$, online serving first reads the latest versioned cache
entry. Consumption-side validation independently repeats protocol and
legality checks before applying any override. If the entry passes, its
pipeline mode, channel subset and quotas, CTR/CVR models, fusion
weights, and reranker settings override the incumbent values;
otherwise, the pipeline executes the incumbent production
configuration. The downstream cascade then returns the recommendation
list and emits traceable execution logs.

The cache is keyed by user and scenario identifiers, and each entry
has a configurable time-to-live (TTL) that governs serving-side
entry expiration. This TTL is distinct from the retention period of
the full-stack execution logs described above. A request in a different
scenario therefore maps to a distinct cache entry rather than causing
the previous entry to expire immediately. Other context changes that
are not encoded in the cache key do not immediately invalidate the
current entry; they affect serving only after the near-line agent
generates and atomically writes a new valid version for the same key,
or after the current entry expires according to its configured TTL.
Concurrent updates for the same key are serialized by the iGraph write
path.

This dual-ended validation---generation-side checks before cache writes
and consumption-side checks before application---prevents malformed,
expired, or protocol-incompatible entries from being applied as
overrides. Generation, validation, or storage failures do not overwrite
the latest valid cache entry. At consumption time, a cache miss, an
expired entry, a read timeout, or failed consumer-side validation
triggers the incumbent configuration for the current request. The
master kill switch bypasses MARS and restores incumbent behavior for
all traffic. Full-stack logs link generated, validated, served,
fallback, and explored configurations to platform-wide impressions,
clicks, cart additions, purchases, and paid amounts by user ID and
timestamp.

\paragraph{Data handling scope.}
MARS operates within Taobao's existing recommendation data infrastructure.
The features entering the prompt are selected from fields already
available to the production recommendation pipeline, and training data
consists of logged request--response--feedback tuples collected under
the same environment. The teacher model (Qwen3.6-Max) is deployed
within the same internal infrastructure and processes the same input
fields. Cross-scenario behavioral sequences are sourced from the platform-wide behavioral logging service, with lookback windows varying by scenario, behavior type, and feature definition. Logging, feature access, and execution-log retention are handled within Taobao's existing recommendation data infrastructure.

\subsection{Post-Training Overview}
\label{sec:post_training}
As shown in the Post-training area of Figure ~\ref{fig:overview}, the agent policy is trained through a progressive three-stage pipeline, each stage initializing the next. Training data combines exposure and exploration logs from the Logging module, while platform-wide feedback—clicks, cart additions, purchases, and paid amounts—provides the reward signal. SFT first transfers schema-valid structured generation from a Qwen3.6-Max teacher and then aligns to recommendation efficiency with Wilson-selected execution evidence. FAPO performs offline preference learning over executed candidates while anchoring full-response generation, restricting preference comparison to differing action values through ActionMask, and weighting pairs by posterior V2P. GRPO samples on-policy configurations, executes them in the snapshot-based Sandbox, scores their lists with a frozen learned reward model, and updates the policy without exposing candidates to users. The updated policy is deployed back to the LLM policy module for subsequent near-line generation. Qwen3-0.6B~\citep{yang2025qwen3} is the near-line agent policy backbone. Sections ~\ref{sec:sft}–~\ref{sec:GRPO} detail each stage.

% %----------------------------------------------------------------------------------------

\section{Problem Formulation}
\label{sec:problem_formulation}
A request is $x = (u, s, c) \in \mathcal{X}$, where $u$ is the user, $s \in
\mathcal{S}$ the scenario and $c$ the context (page number, timestamp,
season, inferred intent); requests follow the online distribution
$\mathcal{D}$.
Conditioned on the request-time serving state, the deterministic operator
$F(x,a)$ returns the ordered item list produced by a valid orchestration $a$
for request $x$. The validation and fallback rules below determine which
orchestration is actually executed. We use $b$ for user feedback and $P$ for
its unknown distribution conditional on the request and displayed list.
\subsection{Optimization Objective}
\label{sec:optimization_objective}
The LLM policy asynchronously emits a response
$y\sim\pi_\theta(\cdot\mid x)$. If parsing succeeds,
$a=\mathrm{Parse}(y)$ is the candidate orchestration. A binary validator
returns $V(a)=1$ when the configuration satisfies the required schema, uses
allowed channels and models, and respects the parameter bounds in
\S\ref{sec:action_space}; otherwise, $V(a)=0$. Let $a_c(x)$ denote the latest valid
cached orchestration, or the incumbent production configuration $a_0(x)$ when
no valid cache exists. The effective action $a_{\mathrm{eff}}(y,x)$ for
subsequent requests is $a$ if parsing, validation, and the cache update succeed,
and $a_c(x)$ otherwise.

Let $\ell=F(x,a_{\mathrm{eff}}(y,x))$ denote the list displayed when the
effective orchestration is served. The \emph{feedback} $b$ records the user's
subsequent behavior, such as clicks, purchases, and paid amounts. A
scenario-specific score $r_s(\ell,b)$ measures business outcomes, and the final
reward is $R_s(a,\ell,b)=r_s(\ell,b)$. We maximize
$J(\theta)=\mathbb{E}_{x,y,b}[R_s(a_{\mathrm{eff}},\ell,b)]$, where
$x\sim\mathcal{D}$, $y\sim\pi_\theta(\cdot\mid x)$, and
$b\sim P(\cdot\mid x,\ell)$. Offline training replaces the business score
with feedback matching or model estimates (\S\ref{sec:reward_design}).
\subsection{Action Space}
\label{sec:action_space}
$F$ exposes seven configurable dimensions, and an \textbf{orchestration} $a$ is
their joint value:
\begin{equation}
a = \big(a_{\mathrm{mode}},\; R,\; q,\; m_{\mathrm{ctr}},\; m_{\mathrm{cvr}},\;
w,\; \eta\big) \in \mathcal{A}.
\end{equation}
Table~\ref{tab:action} details the space; CTR and CVR denote click-through rate
and conversion rate, respectively.

\begin{table*}[t]
\caption{The orchestration action space. The pipeline modes are \textsc{Full}
(retrieval, pre-ranking, ranking, re-ranking), \textsc{Streamline}
(\textsc{Full} without pre-ranking) and \textsc{Minimal} (retrieval and ranking
only).}
\label{tab:action}
\centering
\small
\begin{tabular}{@{}llp{9.5cm}@{}}
\toprule
Symbol & Meaning & Range \\
\midrule
$a_{\mathrm{mode}}$ & pipeline mode &
\textsc{Full}, \textsc{Streamline} or \textsc{Minimal} \\
$R$ & retrieval channel subset &
$R \subseteq \mathcal{R}$ with $3 \le \lvert R \rvert \le 7$ \\
$q$ & per-channel quota &
$q_i = q_i^{\mathrm{base}}(s)\,(1 + \delta_i)$ with $\delta_i \in [-0.2, 0.2]$ \\
$m_{\mathrm{ctr}}, m_{\mathrm{cvr}}$ & CTR / CVR model &
the models available in scenario $s$ \\
$w$ & per-stage fusion weights &
$w \in \mathcal{W}^{d}(a_{\mathrm{mode}})$ with $\mathcal{W} =
\{v_{\mathrm{default}} + 0.05\,k : k = -6, \dots, 6\}$ \\
$\eta$ & re-ranker and its weights &
the re-ranking methods available in scenario $s$ \\
\bottomrule
\end{tabular}
\end{table*}

% %----------------------------------------------------------------------------------------

\clearpage
\section{Method}
\label{sec:method}
All three formal training stages use the same prompt and, in the current
system, the same constrained output protocol. Each stage directly targets
one recommendation-specific challenge: cold-start structured generation
and recommendation-efficiency alignment, format-anchored off-policy
preference learning, and sandbox-based on-policy reinforcement learning.

\begin{figure*}[t]
\centering
\includegraphics[width=\textwidth]{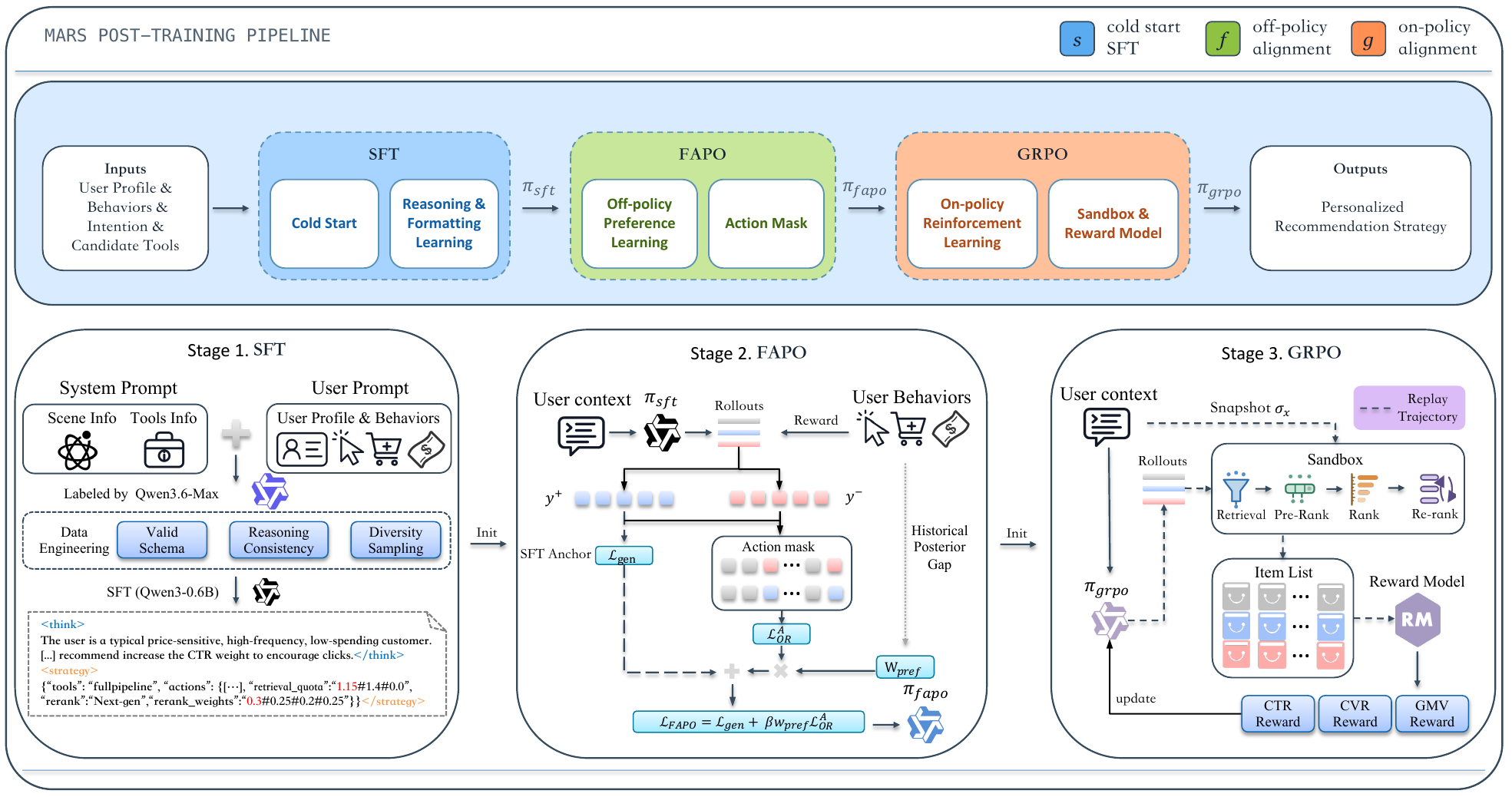}
\caption{Post-training pipeline: schema-aligned SFT, off-policy FAPO with posterior weighting and ActionMask, and on-policy GRPO with Sandbox execution and learned reward estimation. (Note: the \texttt{<strategy>} snippet in Stage~1 uses a schematic shorthand; see Appendix~\ref{app:action-example} for the full production wire format.)}
\label{fig:posttrain}
\end{figure*}
\subsection{Prompt Design and Expert-Knowledge Injection}
\label{sec:prompt_design}
At the system level, the prompt states the orchestration task, scenario context,
formal action space, model performance, parameter bounds, and required response
schema.  It defines Full, Streamline, and Minimal pipeline modes.
At the user level, inputs include profile attributes;
aggregated order counts, GMV, units purchased, and active days over multiple
windows; cross-scenario behavior sequences recording clicks, purchases, and
dwell times; inferred intent; search history; and memory features.
These signals help the model understand the user's current visit intent and
interests, informing pipeline selection and strategy adjustments across
recommendation stages.
In particular, user-behavior and contextual signals allow the agent to assess promotion-abuse and pseudo-exposure tendencies---where a surface is displayed but the user's actual goal lies elsewhere (e.g., a pass-through visit on the way into another scenario)---which inform its choice of pipeline mode.

Appendix~\ref{app:prompt_details} presents the two-message prompt
structure: a publication excerpt of the system prompt---covering system instructions,
expert knowledge, and action constraints---and a fully synthetic user-prompt
example with request-specific context.
This supplementary material elaborates the inputs described above without
altering the formal action space defined in Section~\ref{sec:action_space}.

\paragraph{Context feature engineering.}
The production recommendation system maintains hundreds of user- and item-side
features, yet the prompt's context window imposes a strict token budget.
Naively injecting the full feature set is impractical for three reasons:
(i)~the serialized representation exceeds the available context length;
(ii)~ID-type features (e.g., hashed user or item identifiers) carry no
semantics that an LLM can interpret or reason about; and
(iii)~low-importance features dilute the informative signals with noise,
degrading inference quality.
We therefore apply an importance-based feature selection step before
prompt assembly. Two complementary methods---XGBoost feature importance
and FSCD (Feature Selection based on feature Complexity and variational
Dropout)~\citep{ma2021towards}---are used to evaluate each candidate
feature's contribution to the downstream CTR and CVR prediction tasks.
The highest-ranked features by aggregated importance are retained and
serialized into the prompt's \emph{model feature importance} block,
so that the prompt prioritizes informative and semantically meaningful
signals within the available token budget.
Predictive feature importance and token-budget selection determine
which features enter the prompt for orchestration quality; they do
not establish privacy necessity or suitability for public disclosure.

\paragraph{Expert-knowledge injection.}
Expert-knowledge injection is a key component of MARS prompt design.
It translates production experience into concrete guidance on configuration
defaults, parameter ranges, and user-specific adjustments.
Without this guidance, a base LLM can generate parameter values outside
the operational ranges of the recommendation system, such as a fusion
weight of 100.
We inject three categories of recommendation-domain expert knowledge into
the system prompt to connect user understanding with executable
orchestration decisions.

\emph{(1)~System default configurations.}
Each scenario provides a set of baseline parameters established through
extensive production tuning.
These defaults provide a practical starting point for personalized
adjustments and are not assumed to be optimal for every individual user.
For example, the default retrieval quotas for the Baiyibutie scenario are
listed in Table~\ref{tab:quota-defaults}, with analogous defaults defined
for every supported scenario.

\emph{(2)~Parameter space discretization and bounds.}
The agent explores parameter adjustments within a bounded neighborhood
of the production defaults.
Retrieval quotas must be positive integers perturbed from their defaults
within a $\pm 20\%$ band (Eq.~\eqref{eq:quota-perturbation}).
Fusion weights at the pre-ranking, ranking, and re-ranking stages are
perturbed in steps of $0.05$ from their stage-specific defaults.
The perturbation level $k \in \{-6, \ldots, 6\}$ defines a 13-point
candidate grid per weight dimension (Eq.~\eqref{eq:weight-discretization});
after applying non-negativity and scenario-specific constraints, the set of
legal values for a given weight may be smaller, so each dimension admits
up to 13~discrete levels.
The weights are \emph{not} required to sum to one; each dimension is
adjusted independently.
This design turns open-ended numerical generation into a structured
choice among parameter values centered on established configurations.
Note that these serving-stage fusion weights, which control how predicted
scores are combined during online ranking
(Eq.~\eqref{eq:fusion}), are distinct from the reward and objective
combination coefficients used during post-training
(Section~\ref{sec:reward_design}).

\emph{(3)~Behavior- and intent-based guidance.}
Domain experts translate recommendation experience into adjustment
guidelines based on observable interactions and current intent.
The guidelines describe request-level behavioral contexts rather than
assigning users to fixed groups.
They are expressed in natural language within the prompt and cover the
following representative situations:
\begin{itemize}[nosep,leftmargin=1.5em]
  \item \textbf{Clear purchase intent.}
    When recent interactions indicate a concrete purchase goal,
    prioritize increasing \texttt{cvrWeight} to support efficient
    purchase decisions.
    Keep retrieval focused on relevant items while retaining some
    exploration, and adjust price-related preferences in line with
    the current shopping context.

  \item \textbf{Exploratory browsing.}
    When browsing activity has not yet developed into a clear purchase
    goal, increase \texttt{ctrWeight} to surface engaging items and
    expand retrieval breadth and depth to explore potential interests.
    Reducing \texttt{priceWeight} can also shift emphasis away from
    higher-priced items during exploration.

  \item \textbf{Promotion-related interactions.}
    When recent activity involves coupons, subsidized items, or
    price-comparison intent, decrease \texttt{priceWeight} to emphasize
    value for money.
    If relevant items have been added to the cart, slightly increase
    \texttt{cvrWeight} to support purchase completion.
    Balance retrieval relevance and exploration around the products
    and offers involved in the current shopping activity.

  \item \textbf{Early-stage browsing.}
    When the user is in an initial exploration phase with few recent
    interactions, increase \texttt{ctrWeight} and moderate
    \texttt{priceWeight} to surface engaging, approachable items.
    Broaden retrieval coverage to gather additional interest signals
    and support discovery.

 \item \textbf{Pre-purchase cart evaluation.}
    When the user has added items to the cart and is actively
    comparing options, increase \texttt{cvrWeight} and focus retrieval
    on closely related items to support comparison and
    decision-making.
    When accompanied by coupon activity, moderate \texttt{priceWeight}
    to highlight affordability; otherwise retain its default value.
    Maintain \texttt{ctrWeight} to preserve content appeal.
\end{itemize}

\noindent
These guidelines provide actionable \emph{directional guidance}---which
weights to raise or lower, by roughly how much, and for what business
purpose.
They serve as contextual priors rather than fixed prescriptions:
multiple behavioral signals may coexist within a request, and the
appropriate adjustment depends on the current intent and scenario.
The model combines this guidance with the request context to select
the final configuration.
Together, the production defaults, discretized parameter space, and
behavior-based guidance provide a practical foundation for personalized
orchestration.

\paragraph{Impact of expert-knowledge injection.}
Table~\ref{tab:expert-knowledge-ablation} compares two system-prompt
variants using the same Qwen3-0.6B checkpoint after format SFT across
two representative production scenarios. The variants differ only
in whether recommendation-domain expert knowledge is included in the
system prompt; neither uses the original pretrained base model
directly. Both variants use the same schema and legality validator
and the same incumbent-fallback mechanism.

In this prompt-ablation experiment, both variants achieved
request-level valid-cache coverage above $99\%$ during the same
evaluation window, so fewer than $1\%$ of requests in either arm
used the incumbent fallback. This limits the prevalence of fallback
in the comparison, but does not by itself establish that fallback
differences made a negligible contribution to the observed
business-metric differences. Format validity and business
effectiveness remain distinct properties: a configuration may be
schema-conformant and operationally legal while still making
ineffective pipeline, model, retrieval, or weight choices.

Without expert knowledge, the relative V2P point estimates are
$-5.44\%$ on Baiyibutie and $-27.00\%$ on Miaosha; the corresponding
GMV estimates are $+0.39\%$ and $-21.23\%$. With expert knowledge,
the V2P and GMV point estimates are $+1.57\%$ and $+1.76\%$ on
Baiyibutie, and $+0.20\%$ and $+2.40\%$ on Miaosha. All four metrics show improvements when expert knowledge is injected, supporting the practical value of domain-informed prompting under the evaluated settings.

\begin{table}[t]
\caption{Online point estimates for expert-knowledge injection.
All entries are relative changes (\%) versus the production control.
Both variants use the same Qwen3-0.6B checkpoint after format SFT
and differ only in whether recommendation-domain expert knowledge
is included in the system prompt.}
\label{tab:expert-knowledge-ablation}
\centering
\small
\setlength{\tabcolsep}{4pt}
\renewcommand{\arraystretch}{1.12}
\begin{tabular*}{0.92\columnwidth}{@{\extracolsep{\fill}}llrr@{}}
\toprule
Scenario & Configuration & V2P (\%) & GMV (\%) \\
\midrule
\multirow{2}{*}{Baiyibutie}
  & w/o expert knowledge & $-5.44$ & $+0.39$ \\
  & w/ expert knowledge  & $+1.57$ & $+1.76$ \\
\addlinespace
\multirow{2}{*}{Miaosha}
  & w/o expert knowledge & $-27.00$ & $-21.23$ \\
  & w/ expert knowledge  & $+0.20$ & $+2.40$ \\
\bottomrule
\end{tabular*}
\end{table}

\subsection{Reward Design}
\label{sec:reward_design}
Before describing each training stage, we present the two reward signals
that they share. The \textbf{formula-based business score}
(Section~\ref{sec:business_score}) provides an offline matching proxy
derived from real user behaviors and is used to select training targets
for SFT (Section~\ref{sec:sft}) and to construct preference pairs for
FAPO (Section~\ref{sec:fapo}). The \textbf{learned reward model}
(Section~\ref{sec:reward_model}) is used to score candidate lists
beyond those observed during exposure and serves as the reward signal
for on-policy exploration in GRPO (Section~\ref{sec:GRPO}).

\subsubsection{Formula-based Business Score}
\label{sec:business_score}

We derive formula-based business scores by matching each executed
orchestration's ranked item list against the user's platform-wide
clicks, cart additions, and purchases over the following two days.

The \textbf{composite business score} additionally incorporates click
and cart-addition matches and combines interaction coverage through hit
rate, ranking quality through normalized discounted cumulative gain
(NDCG), and matched transaction value (GMV):
\begin{equation}
S_{\mathrm{comp}}(a)
=
w_1^{\mathrm{comp}}\,\mathrm{HitRate@}K
+
w_2^{\mathrm{comp}}\,\mathrm{NDCG@}K
+
w_3^{\mathrm{comp}}\,\mathrm{GMV}.
\label{eq:composite_score}
\end{equation}
Unless otherwise specified, we use
\begin{equation}
K=10,
\qquad
\left(
w_1^{\mathrm{comp}},
w_2^{\mathrm{comp}},
w_3^{\mathrm{comp}}
\right)
=
\left(
0.75,0.15,0.10
\right).
\end{equation}

At cutoff $K=10$, \textbf{HitRate} measures coverage of the user's
subsequent clicks, cart additions, and purchases in the top-$10$
positions at item, identical-product, and category granularities.
\textbf{NDCG} uses these behaviors as relevance labels, applying
exponential relevance gains and logarithmic rank discounts normalized
by the ideal ordering.

The \textbf{GMV} term in Eq.~\eqref{eq:composite_score} sums realized
transaction amounts only for purchased items present in the candidate
list within the same two-day window (rather than item display prices
or the reward model's expected transaction amount $z_{\mathrm{GMV}}$ in
Section~\ref{sec:reward_model}), and maps the sum to a bounded scale.
Concretely, let $m_i(a)$ indicate whether item $i$ in the list
$\mathcal{L}(a)$ produced by orchestration $a$ matches a purchase in
the user's subsequent two-day platform-wide behavior log, and let $M_i$
denote the realized transaction amount of that purchase. We first sum
the matched transaction amounts:
\begin{equation}
\mathrm{GMV}_{\mathrm{raw}}(a)
=
\sum_{i\in\mathcal{L}(a)}
m_i(a)\,M_i.
\label{eq:formula-gmv-raw}
\end{equation}
To align its numerical scale with $\mathrm{HitRate@}10$ and
$\mathrm{NDCG@}10$, we apply a fixed empirical percentile map
$P_{\mathrm{amt}}(x) = \widehat{F}_{\mathrm{GMV}^{+}}(x) \in [0, 1]$,
where $\widehat{F}_{\mathrm{GMV}^{+}}$ is the empirical cumulative
distribution of positive $\mathrm{GMV}_{\mathrm{raw}}$ values in the
reference production logs (estimated via linear interpolation between
quantile anchors, with $P_{\mathrm{amt}}(0)=0$ and out-of-range values
clipped to $[0, 1]$). The resulting $\mathrm{GMV}$ component in
Eq.~\eqref{eq:composite_score} is
\begin{equation}
\mathrm{GMV}
=
P_{\mathrm{amt}}
\left(
\mathrm{GMV}_{\mathrm{raw}}(a)
\right)
\in [0, 1].
\label{eq:formula-gmv}
\end{equation}

For SFT and FAPO, execution produces multiple lists per request, but
only one is exposed to the user. Both formula-based business scores use
the same subsequent two-day platform-wide behavior log; they are
\emph{offline matching proxies} rather than causal responses to each
list's exposure. The \emph{purchase-only} business score restricts
behavioral matching to purchase events within the same two-day window,
providing a purchase-oriented matching signal for preference-pair
construction in Section~\ref{sec:fapo}.

\subsubsection{Learned Reward Model}
\label{sec:reward_model}

Formula-based matching can score any replayed list when matured
behavioral labels are available, but sparse feedback often yields zero
or tied scores that provide insufficient gradient for policy
optimization. This limitation is most evident in the GRPO stage
(Section~\ref{sec:GRPO}), where the agent explores novel orchestrations
in a Sandbox environment without user exposure. We therefore introduce
a learned reward model to produce denser, more graded business signals.
As shown in Table~\ref{tab:rm}, formula-based matching produces zero
scores for $26.13\%$ of sampled orchestrations and leaves all
candidates tied in $52.46\%$ of requests, whereas the learned reward
model reduces these figures to $0.06\%$ and $0.00015\%$, respectively.

\begin{table}[t]
\caption{Non-discriminative business scores (lower is better). Zeros are
counted over sampled orchestrations; ties over requests.}
\label{tab:rm}
\centering
\small
\begin{tabular}{@{}p{4.0cm}rr@{}}
\toprule
Failure case & Formula-based & Reward model \\
\midrule
Zero score (orchestrations) & $\mathbf{26.13\%}$ & $\mathbf{0.06\%}$ \\
All tied in beam (requests) & $\mathbf{52.46\%}$ &
$\mathbf{0.00015\%}$ \\
\bottomrule
\end{tabular}
\end{table}

We use a decoder-only reward model $\hat{r}_\psi$ to estimate the
user's expected response to a recommendation list. The backbone jointly
processes user and scenario features, the user's historical click
sequence, the final recommendation list produced by the sampled
orchestration, and the ranker's CTR/CVR scores. Each recommended item's
hidden representation feeds separate CTR and CVR prediction heads. CTR
estimates the click probability given an impression, whereas CVR
estimates the purchase probability conditional on a click, both
conditioned on the request and recommendation-list context.

Training optimizes only the click and conversion prediction tasks using
binary cross-entropy. The CTR loss uses impression-level click labels,
whereas the CVR loss is applied only to clicked samples; unclicked
impressions do not contribute to the CVR loss. Training uses a rolling
window of the most recent two months of production logs; the model is
refreshed daily as a routine job. Once trained, $\hat{r}_\psi$ is kept
frozen throughout each GRPO optimization run. The model's area under
the receiver operating characteristic curve (AUC) is $0.76$ for CTR and
$0.81$ for CVR, evaluated on held-out data from a later time window
within the same scenario.

GMV is not a separate prediction target; we derive three list-level
business-score components from the predicted probabilities:
\begin{equation}
\label{eq:rmcomponents}
\begin{aligned}
z_{\mathrm{CTR}}
&=
\sum_i \hat{y}_{\mathrm{ctr}}(i), \\
z_{\mathrm{CTCVR}}
&=
\sum_i
\hat{y}_{\mathrm{ctr}}(i)
\hat{y}_{\mathrm{cvr}}(i), \\
z_{\mathrm{GMV}}
&=
\sum_i
\hat{y}_{\mathrm{ctr}}(i)
\hat{y}_{\mathrm{cvr}}(i)\,p_i.
\end{aligned}
\end{equation}
Here, $p_i$ is the item display price, and
$\hat{y}_{\mathrm{ctr}}(i)\hat{y}_{\mathrm{cvr}}(i)$ estimates the
click-through conversion rate (CTCVR), the joint probability of a click
and subsequent purchase. Thus, $z_{\mathrm{CTCVR}}$ is the expected
number of click-through purchases, and $z_{\mathrm{GMV}}$ is the
expected transaction amount (distinct from the formula-based matched
$\mathrm{GMV}$ in Eq.~\eqref{eq:formula-gmv}, which sums realized
transaction amounts from logged purchases).

To align component scales, we fit empirical percentile maps $P_k$ using
only the reward-model training split and set
\begin{equation}
q_k=P_k(z_k)\in[0,1],
\qquad
k\in
\left\{
\mathrm{CTR},
\mathrm{CTCVR},
\mathrm{GMV}
\right\},
\label{eq:rm-component-normalization}
\end{equation}
clipping values outside the observed training range to $0$ or $1$.
Scenario-specific weights then combine these business components:
\begin{equation}
\label{eq:rmscore}
S_{\mathrm{RM}}(a)
=
w_1^{\mathrm{RM}}\,q_{\mathrm{CTR}}
+
w_2^{\mathrm{RM}}\,q_{\mathrm{CTCVR}}
+
w_3^{\mathrm{RM}}\,q_{\mathrm{GMV}}.
\end{equation}
The coefficients $w_1^{\mathrm{RM}}$, $w_2^{\mathrm{RM}}$, and
$w_3^{\mathrm{RM}}$ are scenario-specific and are distinct from the
composite-score weights in Eq.~\eqref{eq:composite_score}.
Section~\ref{sec:GRPO} describes how $S_{\mathrm{RM}}$ integrates with
the on-policy optimization loop.

\subsection{SFT}
\label{sec:sft}
To satisfy strict near-line throughput and compute budgets, we use the lightweight
Qwen3-0.6B model~\citep{yang2025qwen3} as the near-line agent policy backbone. We train it with
an SFT curriculum with two substages that first establishes structured
generation and then incorporates execution-grounded recommendation knowledge.
\subsubsection{SFT substage 1: cold-start teacher distillation}
\label{sec:sft_substage1}
Direct prompting is insufficient for a $0.6$B model to reliably produce
complex, schema-compliant orchestrations. We therefore use Qwen3.6-Max as a
teacher to generate request-conditioned responses containing explicit reasoning
traces and structured orchestrations from the prompt in \S\ref{sec:prompt_design}.
The teacher operates within the same internal infrastructure as the
production pipeline; its deployment environment and data-access scope
are governed by existing platform policies not detailed in this report.
Rejection sampling removes outputs that violate the serving protocol or schema,
or exhibit inconsistent reasoning and actions; diversity sampling then balances
task types and parameter combinations. Supervised training on the resulting
dataset transfers these structured generation patterns to the student and
raises its format-validity rate from $1.0\%$ to $99.8\%$.
\subsubsection{SFT substage 2: recommendation-efficiency alignment via Wilson selection}
\label{sec:sft_substage2}
The model from SFT substage~1 produces diverse, schema-valid responses, but teacher
supervision alone does not reveal how an orchestration affects the downstream
recommendation system. We therefore continue SFT on execution-verified targets
selected with a Wilson lower bound.

For each request, controlled perturbations of the incumbent configuration
produce candidate orchestrations, which are executed and scored as in
\S\ref{sec:reward_design}. Because business-score scales vary across requests, we compare only
candidates evaluated for the same request. For each user $u$ and orchestration
configuration $a$, we aggregate its pairwise wins, ties, and losses over requests in which it
was evaluated. Let $w_{u,a}$, $t_{u,a}$, and $l_{u,a}$ denote these counts,
$n_{u,a}=w_{u,a}+t_{u,a}+l_{u,a}$, and
$\hat p_{u,a}=(w_{u,a}+t_{u,a}/2)/n_{u,a}$. We rank configurations by the Wilson
lower bound~\citep{wilson1927probable}:
\begin{equation}
\label{eq:wilson}
\mathrm{LB}(u,a)
= \frac{
\hat p_{u,a} + \tfrac{z^2}{2n_{u,a}}
- z\sqrt{\tfrac{\hat p_{u,a}(1-\hat p_{u,a})}{n_{u,a}}
+ \tfrac{z^2}{4n_{u,a}^2}}
}{1 + \tfrac{z^2}{n_{u,a}}},
\end{equation}
where $z$ is the standard-normal quantile for confidence level $1-\alpha$.
We use $z=1.96$.
Let $\mathcal{A}_u$ be the configurations evaluated for user $u$. We select
$a_u^+=\arg\max_{a\in\mathcal{A}_u}\mathrm{LB}(u,a)$ and retain the
request-conditioned responses that instantiate $a_u^+$ as positive SFT targets.
This lower-bound ranking suppresses configurations whose high empirical win
rates are supported by few comparisons. Because pairwise outcomes can share
requests and feedback, we use the Wilson score only as a conservative ranking
heuristic and do not claim nominal confidence coverage. Continuing SFT on these
targets grounds the model from SFT substage~1 in observed recommendation efficiency before
preference and policy optimization.

\subsection{FAPO (Format Anchored Preference Optimization)}
\label{sec:fapo}
In our experiments, DPO~\citep{rafailov2023direct} was prone to format collapse,
whereas ORPO~\citep{hong2024orpo} preserved the output format through its
generation loss but diluted action-level differences with shared structural
content; moreover, sparse purchase feedback makes many preferences derived
from business scores unreliable. FAPO retains full-response generation supervision,
introduces ActionMask to counter dilution, and addresses sparse feedback
through purchase-oriented pair construction and posterior-guided weighting.
The policy is initialized from the SFT checkpoint of \S\ref{sec:sft}.
\subsubsection{Preference pair construction}
\label{sec:preference_pair_construction}
For each request, we generate candidate orchestrations from the SFT
model via beam search, execute them through the online serving pipeline
without exposing the resulting lists to users, and score the lists as
described in \S\ref{sec:reward_design}.

The purchase-only business score retains only purchase-related
contributions:
\begin{equation}
\label{eq:purchase_only}
S_{\mathrm{pay}}(y\mid x)
= \lambda_N\,\mathrm{NDCG}_{\mathrm{pay}}@K
+ \lambda_H\sum_{g\in\mathcal{G}}a_g\,
\mathrm{HitRate}_{\mathrm{pay}}^{g}@K,
\end{equation}
where $\lambda_N$ and $\lambda_H$ are non-negative weighting
coefficients for the NDCG and hit-rate components respectively,
$a_g$ is the per-group weight for matching group~$g$, and
$\mathcal{G}$ comprises item, category, and same-product matching.
These coefficients are scenario-specific hyperparameters held fixed
throughout post-training.
Both the NDCG relevance and hit-rate components use purchase signals,
excluding click and cart contributions.

Pairs are constructed within each request under five rules:
(i)~candidates are compared by the purchase-only business score,
falling back to the composite business score when neither candidate
matches a purchase in the scoring window;
(ii)~pairs follow a reward-gap curriculum from large-gap comparisons
to small-gap comparisons;
(iii)~scenario-specific minimum business-score gaps are used;
(iv)~the number of pairs per frequent orchestration configuration is
capped; and
(v)~weak-evidence pairs, defined as composite-fallback comparisons or
comparisons with a small purchase-only business-score gap, are oriented
toward the candidate with higher historical configuration-level V2P
when both estimates are available and their difference satisfies the
relative-gap criterion in \S\ref{sec:posterior_guided}.
Strong-evidence pairs retain their business-score ordering.
Weak-evidence pairs with a missing estimate or an insufficient V2P
gap also retain their business-score ordering.

Here, ``strong'' and ``weak'' refer to the business-score rules, not
to statistical significance or causal identification. Rule~(v)
introduces an observational purchase-oriented heuristic for
weak-evidence pairs rather than establishing that the reoriented
candidate would improve conversion for the current request.
This yields
$\mathcal{D}_{\mathrm{pref}}=\{(x,y^+,y^-)\}$,
where $x$ is the user context and $y^+$ and $y^-$ are the chosen and
rejected responses.
\clearpage
\subsubsection{Posterior-guided preference weighting}
\label{sec:posterior_guided}
The reported procedure constructs a configuration-level V2P map from
three days of production logs and keeps it fixed during training.
Let $p_s(y)$ denote the historical V2P estimate of the orchestration encoded by $y$ in scenario $s$, keyed by its complete cross-stage configuration (retrieval selection, stage models, and fusion weights), and let $\bar{p}_s$ denote the scenario mean. These configuration-level estimates are observational statistics and
may reflect differences in user composition, purchase intent, traffic
allocation, time, and serving conditions across configurations.

Rule~(v) orients a weak-evidence pair toward the candidate with higher
historical V2P only when both estimates are available and
$|p_s(y^+)-p_s(y^-)|\ge\kappa_s\bar{p}_s$,
where $\kappa_s$ is a scenario-specific relative threshold.
This is a relative point-estimate gap criterion, not a statistical
significance test. In our implementation, we set $\kappa_s=0.05$ for Baiyibutie and $\kappa_s=0.15$ for Miaosha (see Section~\ref{sec:experiment-setup}), while internal logging schemas and proprietary traffic statistics are omitted for confidentiality.

Since strong-evidence pairs may still disagree with the historical
V2P estimates, we further weight the preference loss by
\begin{equation}
\label{eq:prefw}
\begin{aligned}
\widehat{\Delta}_{\mathrm{post}}
&= \frac{p_s(y^+)-p_s(y^-)}{\bar{p}_s},\\
w_{\mathrm{pref}}
&= \operatorname{clip}\big(
1+2\widehat{\Delta}_{\mathrm{post}},\,0.2,\,3.0
\big).
\end{aligned}
\end{equation}
Pairs whose chosen response has the lower historical V2P estimate,
which can remain after rule~(v), receive $w_{\mathrm{pref}}<1$ and are
down-weighted rather than reversed. Clipping limits the numerical
influence of these estimates, but it does not correct selection bias
or confounding and does not establish a causal conversion effect.
Pairs missing either estimate receive $w_{\mathrm{pref}}=1$.
\subsubsection{ActionMask}
\label{sec:actionMask}
In MARS, the values of action fields such as \texttt{retrieval}, \texttt{recall\_weights}, \texttt{retrieval\_quota}, and \texttt{rerank\_weights} account for only about $10\%$ of the tokens in each response, while most remaining tokens are shared by $y^+$ and $y^-$. Sequence-level preference objectives that average log-probabilities over the full response, such as ORPO, distribute the score across many shared structural tokens, diluting the relative contribution of action-value differences.

To address this, we introduce \textbf{ActionMask}. For each pair $(y^+,y^-)$, we compare the two orchestrations field by field. For every action field in the decision space whose values differ, we locate its entire value span in both responses, map the span to token positions via subsequence matching, and mark these positions. All other positions, including shared action values and structural tokens, are excluded from the odds-ratio term, while the generation loss still supervises the complete chosen response. Let $\mathcal{A}(y)$ denote the marked positions of $y$ for the current pair. The action-restricted mean log-probability and its corresponding odds are
\begin{equation}
\label{eq:actionmask}
\begin{aligned}
g_{\theta}^{\mathcal{A}}(y\mid x) &= \frac{1}{|\mathcal{A}(y)|}\sum_{t\in\mathcal{A}(y)}\log\pi_{\theta}(y_t\mid x,y_{<t}),\\
o_{\theta}^{\mathcal{A}}(y\mid x) &= \frac{\exp\!\big(g_{\theta}^{\mathcal{A}}(y\mid x)\big)}{1-\exp\!\big(g_{\theta}^{\mathcal{A}}(y\mid x)\big)}.
\end{aligned}
\end{equation}
\clearpage
\subsubsection{Full objective}
\label{sec:full_objective}
FAPO combines a full-response generation loss with the action-restricted odds-ratio preference loss:
\begin{equation}
\label{eq:fapo}
\begin{aligned}
\mathcal{L}^{\mathrm{gen}}(y^+\mid x) &= -\frac{\log\pi_{\theta}(y^+\mid x)}{|y^+|},\\
\mathcal{L}_{\mathrm{FAPO}} &= \mathbb{E}_{(x,y^+,y^-)\sim\mathcal{D}_{\mathrm{pref}}}\!\left[\mathcal{L}^{\mathrm{gen}}(y^+\mid x)-\beta\,w_{\mathrm{pref}}\log\sigma\!\left(\log\frac{o_{\theta}^{\mathcal{A}}(y^+\mid x)}{o_{\theta}^{\mathcal{A}}(y^-\mid x)}\right)\right],
\end{aligned}
\end{equation}
where $\sigma$ is the sigmoid function and $\beta$ balances generation and preference learning. Only the preference term is scaled by $w_{\mathrm{pref}}$, leaving the coefficient of the generation anchor unchanged. ActionMask determines where preferences are compared, whereas $w_{\mathrm{pref}}$ determines how strongly each pair contributes to preference learning.

\subsection{GRPO: Sandbox-based on-policy orchestration search}
\label{sec:GRPO}
FAPO learns from a fixed preference dataset, whereas GRPO samples
fresh orchestrations from the current rollout policy, including
configurations without logged execution lists. Here, on-policy refers
to candidate generation rather than live user interaction.

\paragraph{Sandbox replay.}
For each persisted request $x$, a snapshot $\sigma_x$ preserves the
request meta-parameters and behavior sequence available at request
time, so that these frozen inputs do not carry post-request
information. The Sandbox reconstructs
this stored request context and executes each candidate orchestration
$a$ to produce $\hat{F}(x,a;\sigma_x)$. Reusing $\sigma_x$ ensures a
consistent request context across candidate rollouts, whereas
non-action serving states (such as underlying candidate pools and
live feature stores) reflect the replay-time environment, providing
approximate execution consistency rather than exact historical
counterfactual equivalence. The Sandbox reconstructs pipeline
execution, while the frozen reward model of
Section~\ref{sec:reward_model} separately estimates user responses to
the resulting recommendation lists without live user exposure
(see Section~\ref{sec:ablations} and Section~7 for further discussion
on replay fidelity).

\paragraph{On-policy optimization.}
For each request $x_i$, we sample a pool of up to 32 candidate
responses from the current rollout policy $\pi_{\theta_{\mathrm{old}}}$
at temperature $T{=}1.5$. After format validation removes
schema-invalid candidates, we select the $G{=}8$ mutually most
dissimilar valid candidates as the training group $\{y^{(i,j)}\}$
(measured by $L_1$ distance on the orchestration action vector). The
selected responses yield orchestrations
$a^{(i,j)}=\mathrm{Parse}(y^{(i,j)})$, which the Sandbox executes
under the shared request snapshot $\sigma_{x_i}$ to produce
$\hat{F}(x_i,a^{(i,j)};\sigma_{x_i})$. The resulting lists are scored
by the frozen reward model via Eq.~\eqref{eq:rmscore} to construct the
standard group-relative advantages $\hat{A}^{(i,j)}$ of
GRPO~\citep{shao2024deepseekmath}. Let $\mathbb{E}_{\mathrm{rollout}}$
denote expectation over sampled requests $x_i$ and the
diversity-filtered training groups obtained from
$\pi_{\theta_{\mathrm{old}}}(\cdot\mid x_i)$. Defining the clipped
ratio $\bar{\rho}_t^{(i,j)} = \operatorname{clip}(\rho_t^{(i,j)},
1-\varepsilon_{\mathrm{low}}, 1+\varepsilon_{\mathrm{high}})$, we
optimize the objective with a Kullback--Leibler (KL) penalty:
\begin{equation}
\label{eq:grpo}
\begin{aligned}
\mathcal{J}(\theta)
={}& \mathbb{E}_{\mathrm{rollout}}\!\Biggl[
\frac{1}{G}\sum_{j=1}^{G}\frac{1}{|y^{(i,j)}|}\sum_t \\
&\quad \min\!\Bigl(\rho_t^{(i,j)}\hat{A}^{(i,j)},
\bar{\rho}_t^{(i,j)}\hat{A}^{(i,j)}\Bigr)\Biggr] \\
&\quad - \beta_{\mathrm{KL}} D_{\mathrm{KL}}\!\bigl(
\pi_\theta\,\|\,\pi_{\mathrm{ref}}\bigr),
\end{aligned}
\end{equation}
where
$\rho_t^{(i,j)}
= \pi_\theta(y_t^{(i,j)}\mid y_{<t}^{(i,j)},x_i)\,/\,
  \pi_{\theta_{\mathrm{old}}}(y_t^{(i,j)}\mid y_{<t}^{(i,j)},x_i)$
is the token-level importance ratio evaluated at the standard
decoding temperature ($T{=}1$; the elevated $T{=}1.5$ applies only
during candidate generation, not during ratio computation),
$\varepsilon_{\mathrm{low}}$ and $\varepsilon_{\mathrm{high}}$ are the
asymmetric clipping bounds, and $\beta_{\mathrm{KL}}$ controls the KL
penalty. Both the policy and $\pi_{\mathrm{ref}}$ are initialized from
the FAPO model of \S\ref{sec:fapo}.

\paragraph{Training stabilization.}
To counter the abrupt entropy drop and in-group candidate
homogenization once the policy converges, we combine high-temperature
over-sampling and diversity filtering with asymmetric clipping
(clip-higher, $\varepsilon_{\mathrm{high}} >
\varepsilon_{\mathrm{low}}$) following DAPO~\citep{yu2025dapo}, leaving
more room for exploring low-probability actions. Because high-temperature generation and subsequent diversity selection alter the sampling distribution, the ratio in Eq.~(14), computed at $T{=}1$, is not an importance correction for the actual candidate-generation procedure. Eq.~(14) does not include separate corrections for these two steps and is used as a diversity-enhanced approximation to the untempered on-policy objective; their effect is not separately quantified in this report.

\subsection{Exploratory Extension:
Feedback-Conditioned Iterative Orchestration}
\label{sec:react}
The preceding post-training stages improve complementary aspects of orchestration generation: SFT establishes schema-valid generation, FAPO learns preferences from executed candidates while preserving the output format, and GRPO enables sandbox-based exploration under learned rewards. Nevertheless, these stages retain a single-turn interface that generates a standalone orchestration from the current context. Although single-turn generation leverages user profiles and historical behaviors, it does not explicitly represent how earlier orchestration decisions performed. Consequently, nearby requests may produce similar configurations, and parameter choices tend to remain concentrated around default settings.

As an exploratory extension, we formulate an iterative paradigm
inspired by ReAct~\citep{yao2022react}, where the LLM reasons over historical configurations and execution feedback to refine subsequent orchestration decisions. Unlike the standard ReAct paradigm that interleaves reasoning and synchronous tool calls within a single request, our loop operates asynchronously across requests and days: each \emph{turn} generates an orchestration configuration served to subsequent requests, and each \emph{observation} corresponds to aggregated execution outcomes received after a feedback delay.

\paragraph{Crowd segmentation.}
To organize population-level statistical priors for expert-knowledge
injection and iterative revision, we partition users along three
transaction-statistic dimensions: annual transaction-amount
band~($V$), transaction-day frequency band~($A$), and historical
purchase-conversion-rate band~($C$).
As summarized in Table~\ref{tab:crowd_segmentation},
$V$~is computed from the user's total consumption amount over the
past year, $A$~from the number of transaction days over the past
three months, and $C$~from the purchase conversion rate over the
past six months.
Binarizing each dimension into upper and lower bands yields eight
statistical crowds (\texttt{A}--\texttt{H}), supplemented by a
ninth fallback crowd (\texttt{O}) for cold-start or unprofiled
users whose features across all three dimensions are unavailable.
Each crowd serves as an indexing key to aggregate historical
execution records and provide directional parameter guidance for
users in the same statistical group.
These groups summarize observed transaction statistics within
specified windows; they do not represent stable personality traits
or actual economic capacity.

\begin{table}[t]
\centering
\caption{Crowd segmentation dimensions and combination definitions.}
\label{tab:crowd_segmentation}
\begin{tabular}{ll}
\toprule
\textbf{Dimension} & \textbf{Underlying Feature} \\
\midrule
Annual transaction-amount band ($V$)             & Total consumption amount over the past year \\
Transaction-day frequency band ($A$)             & Number of transaction days in the past 3 months \\
Historical purchase-conversion-rate band ($C$)   & Purchase conversion rate in the past 6 months \\
\midrule
\textbf{Crowd} & \textbf{Statistical Band
  (Amount / Frequency / Conversion)} \\
\midrule
\texttt{A} & Upper $\cdot$ Upper $\cdot$ Upper \\
\texttt{B} & Upper $\cdot$ Upper $\cdot$ Lower \\
\texttt{C} & Upper $\cdot$ Lower $\cdot$ Upper \\
\texttt{D} & Upper $\cdot$ Lower $\cdot$ Lower \\
\texttt{E} & Lower $\cdot$ Upper $\cdot$ Upper \\
\texttt{F} & Lower $\cdot$ Upper $\cdot$ Lower \\
\texttt{G} & Lower $\cdot$ Lower $\cdot$ Upper \\
\texttt{H} & Lower $\cdot$ Lower $\cdot$ Lower \\
\texttt{O} & Fallback (all three features unavailable) \\
\bottomrule
\end{tabular}
\end{table}

\paragraph{Crowds wiki structure and semantics.}
Building on this taxonomy, we construct a \emph{crowds wiki}, organized as a structured JSON dictionary keyed by crowd identifiers. For each crowd within a given scenario, it stores posterior statistical preferences over specific pipeline elements derived from historical execution logs under different orchestration modes. Rather than prescribing complete configurations or fixed parameter values, each entry specifies a target element and a directional adjustment (\texttt{higher} or \texttt{lower}). For example, in the Miaosha scenario, the wiki entry for Crowd~\texttt{F} (lower amount-band $\cdot$
upper frequency-band $\cdot$ lower conversion-band) is represented
as follows:

\begin{small}
\begin{verbatim}
{
  "F": {
    "preferences": [
      {"element": "recall.Rhot_recall", "prefer": "lower"},
      {"element": "recall.i2i_recall",  "prefer": "lower"},
      {"element": "rerank.w1_ctr",      "prefer": "higher"},
      ...,
      {"element": "rank.w2_cvr",        "prefer": "higher"},
      {"element": "rerank.w3_gmv",      "prefer": "lower"}
    ]
  }
}
\end{verbatim}
\end{small}

\noindent This entry encodes two directional adjustments for
pipeline orchestration.  First, it recommends moderating the quotas
of specific retrieval channels (such as \texttt{Rhot\_recall} and
\texttt{i2i\_recall}) for users in this statistical group.  Second,
it suggests increasing the fusion weights for CTR
(\texttt{rerank.w1\_ctr}) and CVR (\texttt{rank.w2\_cvr}) while
moderating the GMV weight (\texttt{rerank.w3\_gmv}) across the
ranking and re-ranking stages.
The directional adjustments are historical statistical priors
derived from aggregated execution logs, not verified individual
preferences or causal effects.

\paragraph{Feedback-conditioned decision making.}
Our iterative decision-making framework combines this crowd-level experience with individual user-level feedback. While the crowds wiki provides shared directional guidance, user-level history complements it by associating a user's earlier configurations with their corresponding efficiency feedback. In the first turn, the agent generates an initial orchestration from the system prompt, user context, crowds wiki, and scenario-default configuration. In subsequent turns, the agent additionally receives the previous action, its observed performance feedback, and any updates in user context before producing a revised configuration. Repeating this process forms an asynchronous \emph{generate--execute--observe--reflect--revise} loop across successive recommendation requests.

\paragraph{Prototype and LLM-based extension.}
To instantiate this loop, our initial prototype selects a historical
orchestration according to its Wilson lower confidence bound over a
recent observation window and applies fixed-step parameter adjustments
guided by the crowds wiki. In the full LLM-based extension, the fixed
update rule is replaced with contextual reflection, allowing the model
to interpret previous outcomes and determine how the configuration
should be revised.

Training and inference use a consistent representation of the initial context and subsequent action--feedback updates. Because interaction history increases the prompt length, context management is applied to retain decision-relevant information while controlling inference cost. The output continues to follow the established structured-action protocol: reasoning is recorded in \texttt{global\_thought}, while executable parameters remain in \texttt{actions}. ReAct therefore expands the decision context available to the agent without altering the downstream execution interface.

\paragraph{Feedback cadence and scope of results.}
In our current engineering design, the crowds wiki is maintained offline in ODPS and synchronized to iGraph through a daily pipeline. Under this daily cadence, observations from earlier days inform subsequent orchestration decisions, although configurations generated within the same day may remain similar; a near-real-time variant would require a streaming pipeline that continuously returns recent execution outcomes. The deterministic prototype establishes the implementation feasibility
of feedback-conditioned revision, whereas evaluation of the full
LLM-based multi-turn extension remains ongoing. Neither the
deterministic prototype nor the LLM-based multi-turn extension is
included in the main online results reported in
Section~\ref{sec:online_results}. Those results evaluate only the
deployed single-turn orchestration policy
(SFT $\to$ FAPO $\to$ GRPO).

%----------------------------------------------------------------------------------------

\section{Experiments}
\subsection{Setup}
\label{sec:experiment-setup}
\textbf{Online evaluation protocol.} Each online A/B comparison runs for two weeks, with user-level stable random assignment into treatment and control buckets, each receiving $5\%$ of
the corresponding scenario's traffic. The production baseline uses the
incumbent scenario-level orchestration. Across the compared arms, the
available recommendation models and their versions are kept consistent; the orchestration policy is the experimental variable. Subsidy policies and budget constraints are kept consistent across arms. Bucket assignment is determined before treatment exposure and remains fixed throughout the experiment. All online A/B metrics use
an intention-to-treat (ITT) analysis over the full assigned treatment and
control buckets, including default-configuration fallback traffic, rather
than only requests with valid cached orchestrations.

\textbf{Deployment and training.} MARS has operated continuously for three
months across Taobao marketing scenarios; we report results for five of them
(Tables~\ref{tab:online} and~\ref{tab:online-ctr}). It serves more than ten
million daily active users. Depending on the scenario, $85\%$--$95\%$ of online
requests use a valid cached orchestration; the remaining $5\%$--$15\%$ fall back
to the incumbent default configuration. The agent uses a Qwen3-0.6B
model~\citep{yang2025qwen3} to meet near-line throughput and compute budgets. The
orchestration policy is retrained once per week. Each run randomly samples on
the order of $10^5$ training examples from logged requests and trains on $32$ H20 GPUs.
The SFT checkpoint initializes FAPO ($\beta=0.5$), whose checkpoint initializes
GRPO. Each stage trains for one epoch (about four hours) using PyTorch AdamW
(learning rate $2\times 10^{-5}$, batch size $32$, $(\beta_1,\beta_2)=(0.9,0.999)$,
$\mathrm{eps}=10^{-8}$, zero weight decay, gradient clipping at $1.0$). 
FAPO constructs the posterior map from three days of production logs and holds it fixed during training, with a weak-evidence score-gap threshold of $0.02$ and $\kappa_s$ set to $0.05$ for Baiyibutie and $0.15$ for Miaosha.
GRPO uses asymmetric clipping
with $\varepsilon_{\mathrm{low}}=0.2$ and $\varepsilon_{\mathrm{high}}=0.28$,
and a KL penalty coefficient $\beta_{\mathrm{KL}}=0.01$. It
samples up to $32$ candidates in parallel per request at temperature 
$T = 1.5$ and, after format validation, diversity-aware over-sampling
retains the $G = 8$ least similar as one training group (\S\ref{sec:GRPO}).

\textbf{Evaluation metrics.}
V2P and uCTR denote the ratios of purchasing and clicking users to
users exposed to recommendations, respectively; pCTR is the
click-to-impression ratio, and GMV is total transaction value.
All reported changes are relative to the production baseline unless
another comparator is explicitly specified in an ablation experiment.
Bracketed nominal 95\% confidence intervals use two constructions:
for the user-level rate metrics (V2P, uCTR) and impression-level pCTR,
a binomial model with the delta method yields an interval on the log
treatment-to-control ratio mapped back to the relative-change scale;
for GMV, we aggregate per-user transaction values by day for each arm
and report a paired $t$ interval over the 14 daily treatment-to-control
ratios. These intervals serve as practical uncertainty summaries; accounting for within-user impression clustering in pCTR or cross-day serial correlation in GMV via block bootstrapping is left to future offline reanalysis.

\subsection{Online Results}
\label{sec:online_results}
\textbf{Scenarios and metric groups.} Three \emph{purchase-oriented surfaces}
use V2P and GMV (Table~\ref{tab:online}): Baiyibutie (``Ten-Billion Subsidy''),
a major Taobao mini-app with platform-subsidized prices; Miaosha (``Flash
Sale''), a time-limited discount channel with high conversion sensitivity;
and ShangouPostPay, the post-payment feed of Taobao's instant-retail service
that converts local-order traffic into incremental far-field e-commerce
purchases. Two \emph{entry and campaign surfaces} use uCTR and pCTR
(Table~\ref{tab:online-ctr}): MarketingBlock, the homepage marketing grid,
and Dacu, covering main and category-specific venues of promotional campaigns.

\begin{table}[t]
\caption{Online V2P and GMV gains over production (with nominal 95\% confidence intervals).}
\label{tab:online}
\centering
\small
\setlength{\tabcolsep}{2pt}
\begin{tabular*}{\columnwidth}{@{\extracolsep{\fill}}lcc@{}}
\toprule
Scenario & V2P (\%) & GMV (\%) \\
\midrule
Baiyibutie           & $+5.22$ $[+4.58, +5.86]$ & $+7.34$ $[+6.18, +8.51]$ \\
Miaosha              & $+3.88$ $[+2.62, +5.15]$ & $+4.83$ $[+4.15, +5.50]$ \\
ShangouPostPay       & $+2.24$ $[+0.10, +4.43]$ & $+7.17$ $[+1.56, +12.77]$ \\
\bottomrule
\end{tabular*}
\end{table}

\begin{table}[t]
\caption{Online click-metric gains over production (with nominal 95\% confidence intervals).}
\label{tab:online-ctr}
\centering
\small
\setlength{\tabcolsep}{2pt}
\begin{tabular*}{\columnwidth}{@{\extracolsep{\fill}}lcc@{}}
\toprule
Scenario & uCTR (\%) & pCTR (\%) \\
\midrule
MarketingBlock & $+1.08$ $[+0.47, +1.70]$ & $+0.99$ $[+0.08, +1.90]$ \\
Dacu           & $+2.27$ $[+1.13, +3.42]$ & $+3.47$ $[+2.24, +4.71]$ \\
\bottomrule
\end{tabular*}
\end{table}

\textbf{Online gains.}
All reported point estimates are positive, and the displayed nominal 95\% confidence intervals have lower bounds above zero; the lower bounds for ShangouPostPay V2P (+0.10\%) and MarketingBlock pCTR (+0.08\%) are close to zero. On the purchase-oriented
scenarios, MARS improves V2P by $+2.24\%$ to $+5.22\%$ and GMV by $+4.83\%$
to $+7.34\%$, led by Baiyibutie (Table~\ref{tab:online}). On the two entry
and campaign surfaces, MARS similarly improves both uCTR and pCTR
(Table~\ref{tab:online-ctr}).

\textbf{Pipeline-mode usage.}
For requests exhibiting promotion-abuse or pseudo-exposure tendencies (such as promotion-focused entry patterns with limited subsequent item-level engagement, or pass-through visits), the agent selects \textsc{Minimal} to save serving resources.

In each scenario, approximately $20\%$ of requests use
\textsc{Minimal} (retrieval and ranking only), $1\%$ use
\textsc{Streamline} (without pre-ranking), and $79\%$ use
\textsc{Full} (all four stages).

Selecting \textsc{Minimal} for these requests bypasses pre-ranking
and re-ranking. The overall deployment with this mode mix is
associated with a $12$\,ms lower mean online serving-path latency
compared with the production baseline.

\subsection{Ablations}
\label{sec:ablations}
% chktex-file 9
We conduct online ablations on Baiyibutie and Miaosha under the protocol in
\S\ref{sec:experiment-setup}, with offline replay used to assess Sandbox
consistency. Due to the computational and traffic costs of two-week
production A/B tests, we report point estimates from single end-to-end
online runs for ablation variants.

\textbf{Training stages.}
Table~\ref{tab:ablation} compares SFT, FAPO, and GRPO against the
same production baseline under the online protocol in
\S\ref{sec:experiment-setup}. GRPO achieves the highest point
estimate for both metrics in both scenarios. On Baiyibutie, the V2P
and GMV gains increase across all three training stages. On Miaosha,
FAPO raises the relative GMV lift from $1.43\%$ to $3.08\%$ while the
V2P lift shifts slightly from $1.77\%$ to $1.66\%$, before GRPO
further improves both metrics. These results support the overall
effectiveness of the progressive post-training pipeline while showing
that gains need not be strictly monotonic for every intermediate-stage
metric.

\begin{table}[t]
\caption{Training-stage relative changes (\%) versus the production baseline.}
\label{tab:ablation}
\centering
\small
\setlength{\tabcolsep}{3pt}
\renewcommand{\arraystretch}{1.12}
\begin{tabular*}{0.90\columnwidth}{@{\extracolsep{\fill}}lrrrr@{}}
\toprule
& \multicolumn{2}{c}{Baiyibutie} & \multicolumn{2}{c}{Miaosha} \\
\cmidrule(lr){2-3}\cmidrule(lr){4-5}
Stage & V2P & GMV & V2P & GMV \\
\midrule
SFT  & $+3.12\%$ & $+4.10\%$ & $+1.77\%$ & $+1.43\%$ \\
FAPO & $+4.80\%$ & $+5.32\%$ & $+1.66\%$ & $+3.08\%$ \\
GRPO & $\mathbf{+5.22\%}$ & $\mathbf{+7.34\%}$ & $\mathbf{+3.88\%}$ & $\mathbf{+4.83\%}$ \\
\bottomrule
\end{tabular*}
\end{table}

\textbf{Sandbox consistency.}
For GRPO, all candidates within a request group reuse the stored
request snapshot $\sigma_x$ (Section~\ref{sec:GRPO}), which fixes
persisted request meta-parameters and behavior sequences. We evaluate
list-level execution repeatability by replaying each fixed request
and orchestration 10 times under the Sandbox setting. For request $x$,
let $L_x^{(b)}$ denote the ordered recommendation list returned by
replay $b$. Over the $\binom{10}{2}=45$ unordered replay pairs, we
define the request-level list-agreement rate as
\begin{equation}
A_x
=
\frac{1}{45}
\sum_{1\leq b<c\leq 10}
\mathbb{I}\!\left\{
L_x^{(b)}=L_x^{(c)}
\right\},
\end{equation}
where equality requires the two replays to return the exact same
ordered list. Across 10{,}000 evaluated requests, the mean request-level agreement rate was $92.5\%$ (median $100\%$). The residual differences come from a time-decay feature computed from the server clock; they only reorder nearly tied items below the top three
and leave the returned item set unchanged, so the primary
list content remains largely stable across replays.

\textbf{Preference objective.}
We compare FAPO against the full-sequence ORPO baseline on Baiyibutie
and Miaosha (Table~\ref{tab:ablation-diff}(a)). FAPO restricts
preference scoring to the complete value spans of differing action
fields through ActionMask and integrates purchase-oriented pair
construction with posterior weighting (\S\ref{sec:fapo}), while
retaining a full-response generation anchor. FAPO achieves higher V2P
and GMV point estimates across both scenarios ($+3.70\%$ V2P and
$+0.32\%$ GMV on Baiyibutie; $+1.38\%$ V2P and $+3.17\%$ GMV on
Miaosha), supporting the combined effectiveness of these components
as an integrated preference objective.

\begin{table}[t]
\caption{Online ablation point-estimate relative differences
(\%, (variant$\,-\,$baseline)\,/\,baseline) comparing
(a)~FAPO vs.\ ORPO and (b)~model-based vs.\ formula-based reward scoring.}
\label{tab:ablation-diff}
\centering
\small
\renewcommand{\arraystretch}{1.08}
\begin{tabular*}{0.90\columnwidth}{@{\extracolsep{\fill}}lrr@{}}
\toprule
Scenario & V2P & GMV \\
\midrule
\multicolumn{3}{@{}l}{\emph{(a) FAPO vs.\ ORPO}} \\
Baiyibutie & $+3.70\%$ & $+0.32\%$ \\
Miaosha & $+1.38\%$ & $+3.17\%$ \\
\midrule
\multicolumn{3}{@{}l}{\emph{(b) Model-based vs.\ formula-based scoring}} \\
Baiyibutie & $+2.30\%$ & $-0.74\%$ \\
Miaosha & $+0.87\%$ & $+0.68\%$ \\
\bottomrule
\end{tabular*}
\end{table}

\textbf{Reward design.}
In offline sampling, the learned reward model substantially reduces
zero scores and all-tied candidate groups compared with formula-based
scoring (Table~\ref{tab:rm}). Online, model-based scoring improves
V2P point estimates by $+2.30\%$ on Baiyibutie and $+0.87\%$ on
Miaosha, while GMV shifts by $-0.74\%$ and $+0.68\%$, respectively
(Table~\ref{tab:ablation-diff}(b)), indicating a metric- and
scenario-dependent trade-off. Because the reward model is trained on
click and conversion labels using binary cross-entropy without direct
transaction-value supervision, its GMV component ($q_{\mathrm{GMV}}$)
relies on a quantile-mapped proxy derived from predicted CTR, CVR,
and display prices, where probability estimation and price-to-GMV
approximation errors may compound.

\subsection{Case Study}
\label{sec:case_studies}
\begin{figure*}[!t]
\centering
\includegraphics[width=\textwidth]{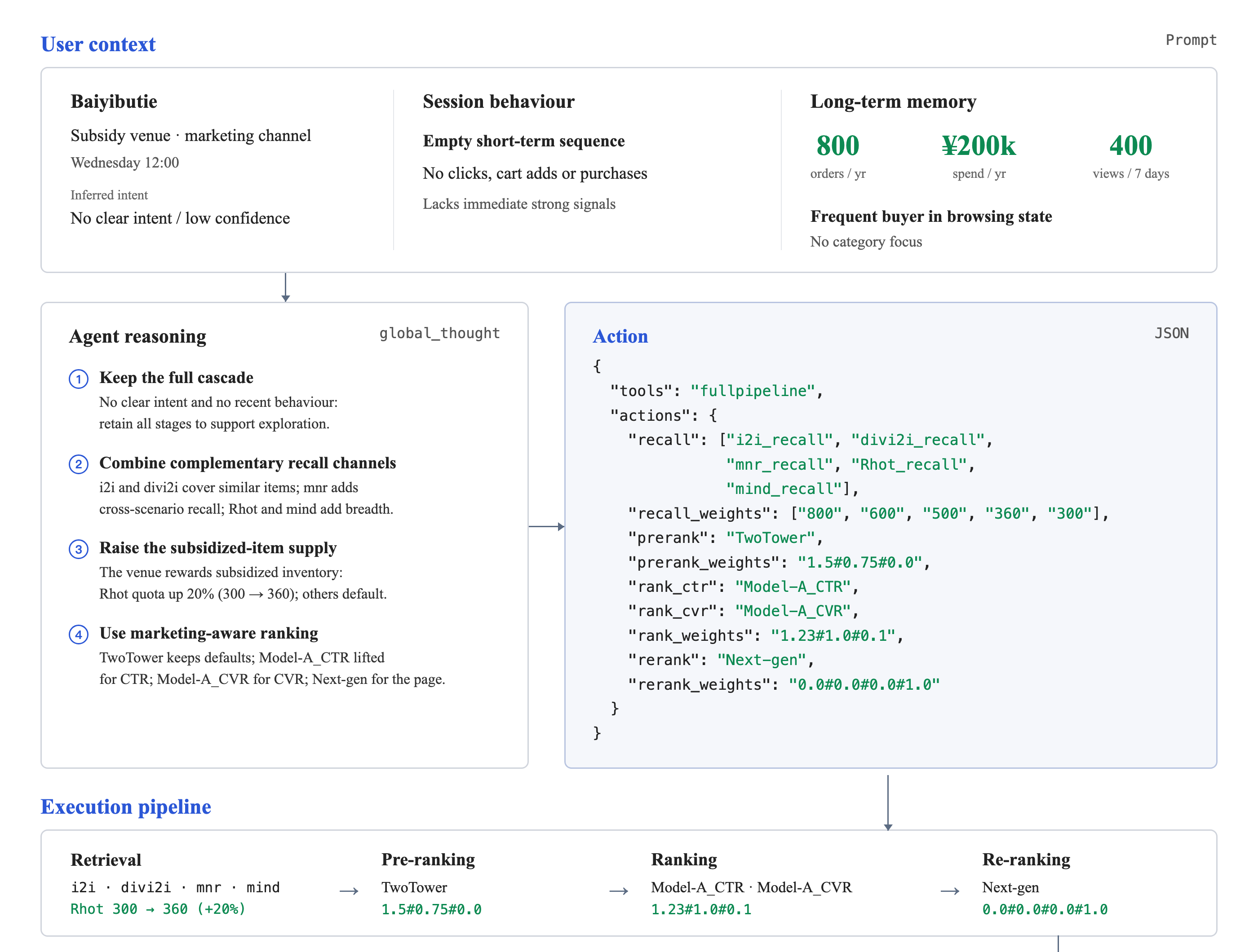}
\caption{A synthetic example of the orchestration workflow in Baiyibutie.
The figure shows synthetic prompt signals (top), the summarized agent
reasoning (middle left), and the grouped JSON action fields (middle right):
the \textsc{Full} pipeline mode, five recall channels with quotas (Rhot
raised to the $+20\%$ bound), the TwoTower pre-ranker, the
publication-labeled Model-A\_CTR and Model-A\_CVR ranking models with
weights $1.23\#1.0\#0.1$, and Next-gen re-ranking with weights
$0.0\#0.0\#0.0\#1.0$. The lower strips show the execution pipeline and an
illustrative item list. Profile values and behavioral statistics are
constructed for illustration and do not reproduce a production user.
The reasoning and action fields are the deployed policy's response to this
synthetic request, with internal identifiers replaced by publication
placeholders.}
% \Description{Case-study figure. The prompt is grouped into scenario context,
% session behaviour, and long-term memory: Baiyibutie with an
% exploratory-browsing label, an empty short-term sequence, approximately
% 200 yearly orders, 25k CNY yearly spend, and 150 views in seven days.
% All profile values are synthetic. Numbered references link these signals to four
% reasoning summaries on the left and grouped JSON fields on the right:
% the full pipeline mode; the recall list i2i, divi2i, mnr, Rhot and mind with
% quotas 800, 600, 500, 360 and 300 (Rhot at the +20\% bound); the TwoTower
% pre-ranker with weights 1.5, 0.75 and 0.0; the publication-labeled
% Model-A_CTR and Model-A_CVR ranking models with weights 1.23, 1.0
% and 0.1; and Next-gen re-ranking with
% weights 0.0, 0.0, 0.0 and 1.0. An execution strip below shows the four stages
% and an illustrative final list of subsidized apparel, hot sneakers, a smart
% device and daily essentials. The final item list is illustrative, not a
% measured outcome; validated orchestrations are written to versioned storage
% and read once per request.}
\label{fig:case}
\end{figure*}

We use a synthetic request to illustrate the end-to-end orchestration
workflow in Baiyibutie (Figure~\ref{fig:case}).  This example demonstrates
the decision and serving interfaces rather than a logged user outcome. The reasoning shown is a summary of the deployed policy's response to the synthetic request, and the action is its parsed output; internal identifiers are replaced with publication placeholders.
The synthetic prompt carries an inferred intent of
\emph{exploratory browsing / low confidence}, an empty short-term behaviour
sequence, and long-term memory showing $800$ yearly orders,
\yen$200$k yearly spend, and $400$ views in the past seven days---a
frequent buyer in a pure-browsing state.
The agent's reasoning connects these signals to concrete decisions.
First, with no clear purchase intent and no recent session activity, the agent retains the full cascade (\textsc{Full}) to preserve broad candidate coverage and support exploration for this request.
Second, frequent purchasing combined with a browsing state
calls for broad candidate coverage: the agent selects five recall channels,
i2i (item-to-item similarity), divi2i (diversity-enhanced similarity), mnr
(multi-scenario near-line retrieval), Rhot (crowd hot), and mind
(multi-interest network with dynamic routing), and raises the Rhot quota to $360$---the
$\delta=+0.2$ upper bound of Table~\ref{tab:action}---to strengthen the
supply of subsidized items while the other channels keep their default
quotas.  Third, because Baiyibutie is a marketing venue, the agent selects
marketing-oriented components: the TwoTower pre-ranker stays at its default
weights $1.5\#0.75\#0.0$, the publication-labeled models
Model-A\_CTR and Model-A\_CVR provide the ranking-stage CTR and CVR
predictions with fusion weights $1.23\#1.0\#0.1$ (the CTR weight two
grid steps above its default), and the Next-gen re-ranker with weights $0.0\#0.0\#0.0\#1.0$ for
sequence-level whole-page optimization; every value belongs to the candidate grid defined in
Eq.~\eqref{eq:weight-discretization} and remains legal after
non-negativity and scenario-specific constraint filtering.  The parsed
orchestration passes validation and is written to versioned storage; for
subsequent requests the online pipeline reads it with one key-value lookup
and executes it.  The bottom strip illustrates a possible item mix of
subsidized apparel, hot sneakers, a smart device, and daily essentials;
these categories are schematic.  This case also shows the structure that
post-training relies on: the reasoning text occupies most of the response
while the decision-relevant values in the \texttt{actions} fields form only
a small fraction, motivating the ActionMask of FAPO (\S\ref{sec:fapo}).

% \begin{figure*}[t]
% \centering
% \includegraphics[width=\textwidth]{Figures/case_study.png}
% \caption{A logged decision for an anonymized high-value browsing user in
% Baiyibutie. Numbered references link prompt signals (top) to summarized
% agent reasoning (middle left) and grouped JSON action fields (middle right):
% the \textsc{Full} pipeline mode, five recall channels with quotas (Rhot
% raised to the $+20\%$ bound), the TwoTower pre-ranker, consguide\_ctr and M2R
% ranking with weights $1.23\#1.0\#0.1$, and Next-gen re-ranking with weights
% $0.0\#0.0\#0.0\#1.0$. The lower strips show the execution pipeline and an
% illustrative item list.}
% \label{fig:case}
% \end{figure*}

%----------------------------------------------------------------------------------------

\section{An Alternative Approach: Traditional Deep Learning}
\label{sec:dnn}
In addition to the LLM agent route, we build a traditional deep
neural network approach for orchestration selection. Rather than
generating a configuration, the DNN scores a predefined set of
candidate orchestrations and selects the highest-scoring one for the
current user context. The two routes share the same input--output
contract, formal orchestration schema, near-line production path, and
downstream execution interface. However, their reachable policy spaces
differ: the LLM generates configurations within the constrained action
space, whereas the DNN can select only from its finite offline recall
pool. In both routes, configurations are produced asynchronously,
validated, and written to a versioned cache before being consumed by
the online cascade. The serving pipeline is therefore unaware of which
route produced a valid configuration, allowing either implementation
to be deployed without modifying the downstream cascade.

The DNN route serves two purposes. First, it provides a discriminative
baseline for evaluating whether a scorer operating over a finite
candidate pool can match the deployed MARS policy. Comparable
performance would indicate that generative inference is not required
under the evaluated candidate coverage and traffic conditions.
Conversely, inferior performance would show that MARS outperforms this
particular tuned DNN implementation; it would not establish that
generative inference is necessary or that all discriminative
approaches are inadequate. Second, the DNN can complement and backstop
the LLM: where it performs comparably, it may absorb a share of
traffic, and it can act as an operational fallback when the LLM is
unavailable. The implementation therefore maximizes reuse of the
existing MARS serving path and uses the same underlying execution logs
and business-score annotations where applicable, while retaining
route-specific candidate construction and optimization objectives.
\subsection{Method}
\label{sec:dnn_method}
\paragraph{Scoring formulation.}
The DNN takes user context $x$ and a candidate orchestration $a$ as
input and produces a pre-sigmoid orchestration logit
\begin{equation}
Q(x,a) = f_{\theta}\big(\phi(x),\, \psi(a)\big),
\label{eq:dnn-score}
\end{equation}
where $\phi(x)$ and $\psi(a)$ denote the user and orchestration
representations. The corresponding bounded pointwise prediction is
$\sigma\!\left(Q(x,a)\right)$, which is trained to fit the bounded
composite business-score target defined below. User features include demographic attributes, purchasing power, and historical behavior sequences shared with the production CTR model. Orchestration features replace item-level features and describe the scenario, candidate identity, pipeline topology, retrieval-channel switches, ranking-model category, and ranking weights. Concretely, the orchestration representation comprises a scenario ID identifying the scenario (e.g., Miaosha, Baiyibutie, or ShangouPostPay); a multi-valued pipeline ID distinguishing Full, Streamline, and Minimal pipelines; one binary switch and one continuous retrieval quota feature per retrieval channel; multi-valued model IDs identifying the model categories used at the pre-ranking, ranking, and re-ranking stages, such as TwoTower, PPNet and Next-gen; and continuous weight features covering the stage fusion weights of pre-ranking, ranking, and re-ranking, including the CTR, CVR, and price weights. Categorical features are embedded, topology and switch features are represented using multi-hot vectors, and retrieval quotas and rank-stage weights are treated as continuous inputs.

\begin{figure*}[!t]
\centering
\includegraphics[width=0.7\linewidth]{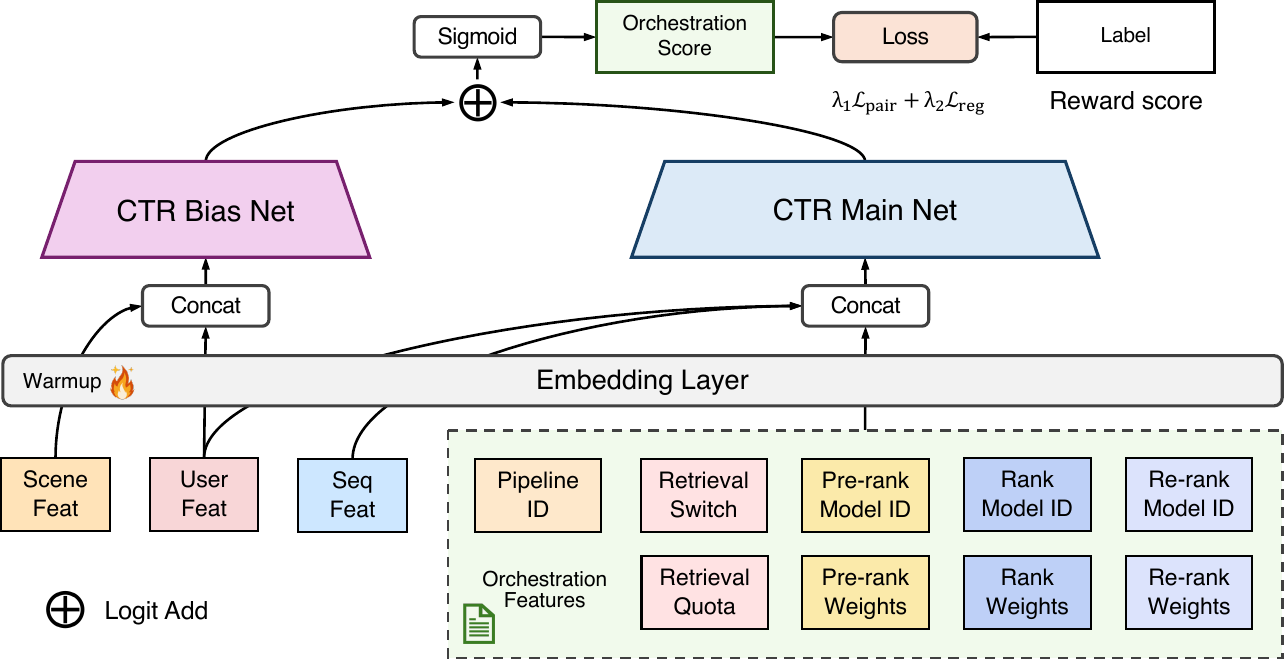}
\caption{Architecture of the DNN orchestration scorer. The backbone and the warm-started embeddings are shared with the production CTR model; only the orchestration features and the scoring head are specific to orchestration selection.}
\label{fig:dnn-arch}
\end{figure*}

\paragraph{Model architecture.}
As illustrated in Figure~\ref{fig:dnn-arch}, the scorer inherits the dual-branch embedding-and-MLP backbone of the production CTR model, comprising a \textit{Bias Net} and a \textit{Main Net}. After passing through a shared embedding layer, scene and user features are concatenated and fed into the Bias Net, while user, sequence, and multi-stage orchestration features are concatenated as inputs to the Main Net. The scalar outputs of the two networks are summed (Logit Add) to
produce the pre-sigmoid orchestration logit $Q(x,a)$. The final bounded pointwise prediction is
$\sigma\big(Q(x,a)\big)$, whereas the pairwise objective operates
directly on differences between logits. The pointwise output estimates the bounded composite business score
of a user--orchestration pair, rather than an item-level click
probability or a calibrated behavioral probability. To ease cold start and inherit existing CTR modeling capacity, parameters associated with overlapping user and sequence features are warmed up from the pretrained production CTR model.

\paragraph{Default-centered spherical exploration as orchestration recall.}
Because the DNN scorer selects the top-scoring configuration at inference, the candidate pool strictly bounds its reachable policy space. MARS constructs this recall set offline via a default-centered spherical exploration strategy. Treating the scenario's incumbent production configuration as the sphere center and the edit distance as the radius, we first discretize the unconstrained combinatorial space. This is achieved by instantiating three standard pipeline templates (\textit{Full}, \textit{Streamline}, and \textit{Minimal}), quantizing continuous parameters, and bounding search ranges to valid production intervals. Each candidate is then assigned a cumulative edit cost: a single weight-step adjustment incurs a cost of 1, whereas switching a model, adding a retrieval channel, or dropping a stage incurs a cost of 3. Using branch-and-bound, we prune candidates whose accumulated cost exceeds 8 for \textit{Full} or 12 for the other templates. In the Miaosha scenario, this pruning eliminates 98.47\% of the
post-discretization candidate set, yielding approximately 59.5K viable
candidates (comprising 46,339 \textit{Full}, 13,008
\textit{Streamline}, and 188 \textit{Minimal} configurations).

To prevent redundant evaluation—as configurations differing by a single weight step yield nearly indistinguishable scores—the recall pool must be further distilled into a compact, high-quality, and mutually diverse subset. We represent each orchestration as a 41-dimensional feature vector and apply a farthest-first greedy traversal. Seeded with the three template defaults, each iteration selects a candidate among the top-5 options that maximize the minimum $L_{1}$ distance to the already-selected set plus $0.3\times$ a domain-prior quality score, with an incremental distance cache ensuring linear time complexity per round. This procedure yields a compact pool of under 1,000 candidates (currently 256, allocated as 176/48/32 across the three templates). On Miaosha, pairwise edit distances within this final set range from 2 to 18 (mean: 11.1). This candidate pool serves as the static DNN recall set between offline refreshes, operating independently of the LLM's beam-search sampling.

\paragraph{Training objective.}
To enable a controlled comparison with the LLM policy, the DNN scorer
reuses the same underlying execution logs and composite business-score
annotations. The optimization objectives nevertheless differ: MARS
uses the staged SFT, FAPO, and GRPO procedure described above, whereas
the DNN is optimized with the pointwise and pairwise objectives defined
here.

Let $S_{\mathrm{comp}}(a)$ denote the composite business score
defined in \S\ref{sec:business_score}, which combines HitRate,
NDCG, and percentile-normalized matched GMV. Under the default
coefficient setting, all three components lie in the unit interval,
and their non-negative coefficients sum to one. Therefore,
\begin{equation}
0
\leq
S_{\mathrm{comp}}(a)
\leq
1.
\label{eq:bounded-composite-score}
\end{equation}
Here, $S_{\mathrm{comp}}(a)$ denotes the component-normalized
composite business score, rather than an unbounded monetary value.

The implemented DNN label pipeline uses this bounded composite
business score directly as the pointwise target:
\begin{equation}
r
=
S_{\mathrm{comp}}(a),
\qquad
0
\leq
r
\leq
1.
\label{eq:dnn-target}
\end{equation}
The DNN target uses neither training-set extrema nor any additional
normalization or clipping. It therefore preserves both the numerical
value and the within-request ordering of the composite business score.

The training objective combines a pointwise fitting loss with an
intra-request pairwise ranking loss:
\begin{align}
\mathcal{L}_{\mathrm{reg}}
&=
\sum_{(x,a)}
\mathrm{BCE}\!\left(
\sigma\big(Q(x,a)\big),\,r
\right),
\label{eq:loss-reg}\\
\mathcal{L}_{\mathrm{pair}}
&=
\sum_{(x,\,a_{\mathrm{win}},\,a_{\mathrm{lose}})}
\operatorname{softplus}\!\left(
-\left(
Q(x,a_{\mathrm{win}})
-
Q(x,a_{\mathrm{lose}})
\right)
\right),
\label{eq:loss-pair}\\
\mathcal{L}
&=
\lambda_{1}\mathcal{L}_{\mathrm{pair}}
+
\lambda_{2}\mathcal{L}_{\mathrm{reg}},
\qquad
\lambda_{1}=\lambda_{2}=1.
\label{eq:loss-total}
\end{align}

Here, $\mathcal{L}_{\mathrm{reg}}$ fits the sigmoid-transformed
prediction $\sigma\!\left(Q(x,a)\right)$ directly to the composite
business-score target $r=S_{\mathrm{comp}}(a)$, with
$0\leq r\leq1$. It is a pointwise score-fitting objective rather than
a behavioral-probability calibration objective.

$\mathcal{L}_{\mathrm{pair}}$ operates directly on logit differences
and encourages the correct relative ordering between superior
$a_{\mathrm{win}}$ and inferior $a_{\mathrm{lose}}$ orchestrations
within the same request. The pairwise term provides
an additional within-request ranking signal that is less sensitive
to variation in absolute reward levels across users and contexts.

% \paragraph{Standalone and hybrid inference.}
% During standalone inference, the DNN scores the candidate configurations and returns the highest-scoring orchestration. Its effectiveness is therefore bounded by candidate coverage: a strong scorer cannot recover an orchestration that is absent from the candidate pool. MARS also supports a hybrid route. The LLM first generates an orchestration, from which several neighboring variants are constructed by changing the pipeline topology, the retrieval quotas, or the ranking model. The DNN scores the original and the derived candidates, and the highest-scoring orchestration is selected. This design separates generative candidate construction from learned value estimation while keeping both routes on the same scoring interface.

\subsection{Experiments}
\label{sec:experiments}
Offline evaluation assesses intra-request ranking performance across
candidate orchestrations. We report two metrics:
\textit{Top-1 Hit Rate}, which measures whether the DNN selects the
candidate with the highest reference composite business score, and
\textit{Win-or-Tie Rate}, which measures whether the DNN-selected
configuration matches or exceeds the reference score of the
LLM-generated orchestration.

Training with the pointwise fitting objective
($\mathcal{L}_{\mathrm{reg}}$) alone yields point estimates of
$0.6220$ for Top-1 Hit Rate and $0.6525$ for Win-or-Tie Rate.
Adding the pairwise ranking objective
($\mathcal{L}_{\mathrm{pair}}$) yields corresponding point estimates
of $0.6224$ and $0.6529$. The observed absolute difference is
$0.0004$ for each metric. Without uncertainty intervals or results
from repeated training runs, these small differences do not establish
a stable advantage for the pairwise objective.

We further compare the deployed MARS LLM policy with the tuned DNN
baseline through concurrent two-week online A/B tests in two core
scenarios, \textit{Baiyibutie} and \textit{Miaosha}. Both routes use
the same executable configuration schema and downstream serving
pipeline, and their training data are derived from the same underlying
execution logs and business-score annotations.

Their candidate construction and optimization procedures nevertheless
differ. The DNN selects from a finite offline recall pool using
pointwise and pairwise objectives, whereas MARS generates
configurations and uses the multi-stage SFT--FAPO--GRPO training
pipeline.

Table~\ref{tab:dnn-online} reports V2P and GMV point-estimate
differences of $+3.36\%$ and $+5.90\%$, respectively, for MARS
versus the tuned DNN on Baiyibutie. The corresponding estimates on
Miaosha are $+2.03\%$ and $+1.76\%$. These observed differences
favour the evaluated MARS policy in both scenarios under the reported
production settings.

\begin{table}[t]
\centering
\caption{Relative online V2P and GMV point-estimate differences
(\%, (MARS$\,-\,$DNN)\,/\,DNN) for the deployed MARS LLM policy
versus the tuned DNN baseline during a concurrent two-week A/B
evaluation across two production scenarios.}
\label{tab:dnn-online}
\begin{tabular}{@{}lcc@{}}
\toprule
Scenario & V2P & GMV \\
\midrule
Baiyibutie & +3.36\% & +5.90\% \\
Miaosha & +2.03\% & +1.76\% \\
\bottomrule
\end{tabular}
\end{table}

We hypothesize that the observed differences may be related to three
architectural distinctions:
(i) generative exploration within the constrained combinatorial
action space versus selection from a finite offline recall pool;
(ii) the LLM's use of semantic representations of user and scenario
contexts; and
(iii) multi-stage SFT--FAPO--GRPO alignment versus
pointwise-pairwise optimization.

% The comparisons above evaluate the standalone DNN and MARS policies rather than the unified hybrid system. End-to-end evaluation of the hybrid routing mechanism, including its latency, compute cost, and business-metric trade-offs, remains future work.

%----------------------------------------------------------------------------------------

\section{Conclusion, Limitations, and Future Directions}
\label{sec:conclusion}
MARS extends recommendation personalization from item-level prediction to the cross-stage configurations that control the recommender pipeline. Instead of applying a single scenario-level configuration to all users, it generates personalized decisions over pipeline modes, retrieval channels and quotas, model choices, fusion weights, and re-ranking parameters. This formulation reuses existing recommendation components while introducing a trainable policy for coordinating them according to user and scenario context.

The system adopts near-line LLM generation to separate orchestration production from online execution. Personalized configurations are generated, validated, and cached before being consumed by the serving pipeline, avoiding synchronous LLM inference on the latency-critical recommendation path. Schema validation, default fallbacks, logging, and rollback mechanisms provide safeguards for production deployment and continuous policy iteration.

The SFT--FAPO--GRPO post-training framework addresses complementary challenges in orchestration generation. SFT establishes schema-valid structured generation and transfers recommendation knowledge into the policy. FAPO learns action preferences from executed candidates while preserving a full-response generation anchor. GRPO further explores candidate configurations through Sandbox execution and learned reward estimation. Together, these stages connect the semantic interpretation of user context with executable decisions across multiple recommendation stages.

Online deployment reports positive business-metric point estimates
across five Taobao scenarios under the evaluated production settings.
The evaluated deployment is also associated with a reported reduction
of $12$\,ms in mean online serving-path latency. These observations
suggest that personalized orchestration can improve recommendation
outcomes while simplifying the serving path. Nevertheless, the current
system remains constrained by the available action space, the fidelity
of offline replay and reward estimation, the difficulty of attributing
aggregate differences to individual serving stages, and the
computational cost of LLM-based configuration generation.

\paragraph*{Expanding the action space.}
The current policy operates within a predefined collection of pipeline modes, model choices, channel switches, quotas, and parameter ranges. These constraints ensure that generated configurations are executable and can be validated before deployment. However, they also bound the policy's optimization capacity. The agent cannot select capabilities that are absent from the action schema, and a useful cross-stage configuration may remain unreachable even when the model correctly understands the user's intent.

Increasing the number of independently adjustable parameters is not sufficient by itself. Decisions at different stages interact with one another: retrieval determines the candidate supply available to ranking, while pipeline simplification changes both computation and the transformations applied to the candidate set. Unstructured expansion could therefore enlarge the search space without improving its effective coverage.

Future work will incrementally introduce additional model choices, finer-grained parameter controls, and further pipeline decisions supported by the underlying system. Each extension should define legal parameter combinations, resource constraints, validation rules, and fallback behavior. The action schema, execution interface, replay environment, and feedback logs must evolve together so that newly introduced actions remain trainable, observable, and safe to deploy.

\paragraph*{Improving replay and reward estimation.}
Sandbox replay enables the evaluation of alternative configurations without exposing every candidate to users. Nevertheless, replay is an approximation of historical serving rather than an exact reconstruction. Persisted request features can reproduce important context, but the available candidate pool, model state, and other serving conditions may change over time. Re-executing a pipeline reconstructs its output under the replay environment; it does not reveal the user response that would have occurred had the alternative list actually been exposed.

Formula-based business scores introduce another boundary. They match candidate lists against subsequent platform-wide behaviors and provide useful offline supervision for SFT and FAPO, but they are matching proxies rather than causal responses to each candidate list. The learned reward model is used to score newly
generated configurations, yet it can inherit noise, bias, and
distribution shift from its training data. These risks become more important as policy optimization explores configurations that differ from historical execution records.

This report does not disclose a minimum per-configuration support-count criterion or the fraction of preference pairs for which both posterior estimates are available. The relative-gap thresholds in Section~\ref{sec:posterior_guided} are not support-count thresholds.

Future work will improve replay coverage, state consistency, and component version tracking. Reward estimation will be evaluated not only through fitting accuracy, but also through calibration, request-local ranking quality, temporal generalization, and agreement with online outcomes. Uncertainty-aware filtering may prevent policy optimization from overusing reward estimates in poorly represented regions of the action space. More reliable replay and feedback could also support ReAct-style iterative revision, although the benefits of multi-turn optimization must be evaluated separately from the additional computation and reward-model error it introduces.

\paragraph*{Isolating serving-stage effects across scenarios.}
Current online experiments evaluate the deployed orchestration policy as an end-to-end system. Consequently, aggregate improvements do not directly identify how much each serving-stage decision contributes. Pipeline selection, retrieval-channel control, quota allocation, model selection, fusion weights, and re-ranking parameters may each affect the final outcome, and their interactions can differ across scenarios and user groups.

Future experiments will isolate these effects through controlled interventions that hold unrelated configuration dimensions fixed where possible. Stage-level ablations should be complemented by selected joint interventions because independently measured effects may not capture cross-stage interactions. Evaluation should examine both business outcomes and serving measurements, including pipeline usage, cache coverage, execution cost, and fallback behavior.

Such analysis should include all assigned requests rather than only those that successfully consume personalized configurations. It should also distinguish end-to-end serving latency from the cost of near-line LLM generation. This will clarify which decisions account for the observed business
point estimates and serving-latency difference, which effects transfer
across scenarios, and where scenario-specific orchestration remains
necessary.

\paragraph*{Reducing inference cost while preserving orchestration quality.}
Near-line generation removes LLM decoding from the synchronous serving path, but it does not eliminate the computation required to generate and refresh personalized configurations. Model size, context length, refresh frequency, and the number of candidates generated during optimization all contribute to the total cost of maintaining the orchestration policy. The observed $12$\,ms serving-latency reduction should therefore be interpreted separately from the resource cost of near-line inference. This serving-path reduction does not indicate faster LLM decoding; decoding is performed near-line, outside the synchronous request path.

Future work will explore model distillation and pruning to reduce this cost. Distillation may transfer the behavior of the post-trained policy to a smaller model, while pruning may remove redundant computation from the existing model. These approaches should not be evaluated solely through schema validity or token-level agreement. A compressed policy must also preserve action quality, useful action-space coverage, robustness to changing user intent, and downstream recommendation performance.

Traditional DNN and hybrid policies provide complementary options. A lightweight discriminative model can score LLM-generated candidates or handle decisions that do not require extensive semantic reasoning. Hybrid deployment may therefore combine the flexibility of generative orchestration with the efficiency of conventional scoring and budget-constrained selection. Determining the appropriate division of responsibility will require matched comparisons of orchestration quality, inference cost, adaptability, and operational complexity.

Overall, future development will focus on maintaining a reliable connection among controllable actions, credible feedback, and measurable online outcomes. Expanding the action space, improving replay and reward estimation, isolating serving-stage effects, and reducing inference cost through distillation and pruning form a coordinated roadmap. The objective is to make personalized orchestration broader and more efficient without compromising decision quality or production reliability.

\newpage

%----------------------------------------------------------------------------------------

% Bibliography components
\phantomsection
\addcontentsline{toc}{section}{References}
\bibliographystyle{abbrvnat}
\nobibliography*
\bibliography{bibtex}

\clearpage

\appendix
\section*{Appendix}
\label{sec:Appendix}
\section{Contributions}

\begin{multicols}{2}
\noindent
\textbf{Core Contributors} \\
Chaoqun Hou \\
Cheng Guo \\
Fengze Jiang \\
Jiaqi Zheng \\
Junhong Lou \\
Longxiang Xiong \\
Lv Shao \\
Ping Liu \\
Qixin Li \\
Tong Liu \\
Xiannian Hu \\
Xinyuan Qi \\
Yong Duan \\
Yu Zhu \\
Zi Zhuang \\
Zixin Zhao \\
Zongwei Wang \\

\noindent
\end{multicols}

\noindent
The listing of authors is in alphabetical order based on their first names.

\newpage

%%%%%%%%%%%%%%%%%%%%%%%%%%%%%%%%%%%%%%%%%%%%%%%%%%%%%%%%%%%%%%%%%%%%%%%%%%%%%%%%%%%%%%%%%%
\section{Notation}

Table~\ref{tab:notation} summarizes the principal symbols used
throughout this report. Several symbols are reused across modules
with distinct meanings; the \emph{Domain / scope} column disambiguates
each occurrence.

{\small
\begin{longtable}{@{}p{3.8cm}p{5.8cm}p{4.2cm}@{}}
\caption{Summary of notation.}\phantomsection\label{tab:notation}\\
\toprule
Symbol & Meaning & Domain / scope \\
\midrule
\endfirsthead

\multicolumn{3}{c}{\tablename\ \thetable\ (continued)}\\
\toprule
Symbol & Meaning & Domain / scope \\
\midrule
\endhead

\midrule
\multicolumn{3}{r}{\emph{Continued on next page}}\\
\endfoot

\bottomrule
\endlastfoot

%% ---- Problem formulation and action space ----
\multicolumn{3}{l}{\emph{Problem formulation and action space}}\\
\addlinespace

$x=(u,s,c)\in\mathcal{X}$ &
  Request: user $u$, scenario $s\!\in\!\mathcal{S}$, context $c$ &
  Global \\

$\mathcal{D}$ &
  Online request distribution &
  Global \\

$y\sim\pi_\theta(\cdot\mid x)$ &
  Full policy-generated response (text) &
  LLM policy; distinct from parsed action $a$ \\

$a\in\mathcal{A}$ &
  Executable orchestration configuration; \newline
  $\mathcal{A}$: constrained seven-dimensional action space &
  Global \\

$\mathrm{Parse}(y)$;\; $V(a)\!\in\!\{0,1\}$ &
  Structured parser; binary schema validator &
  Serving pipeline \\

$a_0(x)$, $a_c(x)$, $a_{\mathrm{eff}}(y,x)$ &
  Incumbent default, latest valid cache, effective action &
  Serving pipeline \\

$F(x,a)$;\; $\ell$ &
  Deterministic pipeline execution; displayed item list &
  Online cascade \\

$b$;\; $P(\cdot\mid x,\ell)$ &
  User feedback; its unknown conditional distribution &
  Global \\

$r_s(\ell,b)$;\; $R_s(a,\ell,b)$ &
  Scenario-specific business score; final reward &
  Global \\

$J(\theta)=\mathbb{E}_{x,y,b}\{R_s\}$ &
  Global online business objective &
  Global; distinct from GRPO objective $\mathcal{J}(\theta)$ \\

$a_{\mathrm{mode}}$ &
  Pipeline mode (\textsc{Full} / \textsc{Streamline} /
  \textsc{Minimal}) &
  Action dim.\ 1 \\

$R\subseteq\mathcal{R}$ &
  Selected retrieval-channel subset &
  Action dim.\ 2 \\

$q_i=q_i^{\mathrm{base}}(s)(1+\delta_i)$, \newline
$-0.2\leq\delta_i\leq0.2$ &
  Per-channel retrieval quota (positive integer) &
  Action dim.\ 3; distinct from percentile $q_k$ \\

$m_{\mathrm{ctr}},\;m_{\mathrm{cvr}}$ &
  Selected CTR / CVR model for the scenario &
  Action dims.\ 4--5 \\

$w\in\mathcal{W}^{d}(a_{\mathrm{mode}})$ &
  Per-stage fusion weight vector; \newline
  $k\!\in\!\{-6,\dots,6\}$, step $0.05$, subject to
  non-negativity and scenario legality &
  Action dim.\ 6; $w$ is also reused with different semantics below \\

$\eta$ &
  Re-ranker selection and its parameters &
  Action dim.\ 7 \\

$\mathrm{score}(i)
=\hat{y}_{\mathrm{ctr}}(i)^{w_{\mathrm{ctr}}}
 \!\cdot\!
 \hat{y}_{\mathrm{cvr}}(i)^{w_{\mathrm{cvr}}}
 \!\cdot\!
 p_i^{w_{\mathrm{price}}}$ &
  Multiplicative item-ranking score &
  Ranking and pre-ranking fusion \\

$w_{\mathrm{ctr}},w_{\mathrm{cvr}},w_{\mathrm{price}}$ &
  Item-score exponents for CTR, CVR, price &
  Ranking / pre-ranking stage \\

$w_{\mathrm{GMV}},w_{\mathrm{rank}}$ &
  Additional Next-gen re-ranker exponents &
  Re-ranking stage \\

\addlinespace
%% ---- Reward design ----
\multicolumn{3}{l}{\emph{Reward design}}\\
\addlinespace

$S_{\mathrm{comp}}(a)$ \newline
$=w_1^{\mathrm{comp}}\,\mathrm{HitRate@}K$ \newline
${}+w_2^{\mathrm{comp}}\,\mathrm{NDCG@}K$ \newline
${}+w_3^{\mathrm{comp}}\,\mathrm{GMV}$ &
  Composite formula-based business score combining hit rate,
  NDCG, and matched transaction value &
  Offline matching \\

$w_i^{\mathrm{comp}},\;w_i^{\mathrm{RM}}$ \newline
$(i=1,2,3)$ &
  Mixing coefficients for the formula-based score and learned
  reward score, respectively; the two sets have different
  semantics and need not share the same values &
  Reward-score aggregation; distinct from serving-stage
  fusion weights \\

$m_i(a)$ &
  Indicator of whether item $i$ produced by orchestration $a$
  matches a purchase in the subsequent behavior window &
  Formula-based reward \\

$M_i$ &
  Realized transaction amount associated with the matched
  purchase of item $i$ &
  Observed purchase data; not item display price \\

$\mathrm{GMV}_{\mathrm{raw}}(a)$ \newline
$=\sum_{i\in\mathcal{L}(a)}m_i(a)M_i$ &
  Sum of realized transaction amounts associated with
  purchases matched by orchestration $a$ &
  Formula-based reward; monetary amount before percentile
  normalization \\

$\widehat{F}_{\mathrm{GMV}^{+}}$ &
  Empirical cumulative distribution of positive action-level
  raw matched-GMV values in the reference production logs &
  Formula-based reward calibration \\

$P_{\mathrm{amt}}(x)
=\widehat{F}_{\mathrm{GMV}^{+}}(x)$ \newline
$0\leq P_{\mathrm{amt}}(x)\leq1$ &
  Fixed empirical percentile map for action-level raw matched GMV;
  $P_{\mathrm{amt}}(0)=0$ &
  Formula-based reward calibration; distinct from the
  reward-model percentile maps $P_k$ \\

$\mathrm{GMV}$ \newline
$=P_{\mathrm{amt}}
 \bigl(\mathrm{GMV}_{\mathrm{raw}}(a)\bigr)$ &
  Matched GMV component in $S_{\mathrm{comp}}(a)$ (realized
  transaction amounts of matched purchases mapped to $[0,1]$) &
  Formula-based reward; $0\leq\mathrm{GMV}\leq1$; distinct from
  online A/B metric $\mathrm{GMV}$ \\

$K$ &
  Recommendation list cutoff &
  Reward evaluation \\

$\hat{r}_\psi$ &
  Learned reward model (decoder-only) &
  GRPO reward signal; $\psi$: reward-model parameters \\

$\hat{y}_{\mathrm{ctr}}(i)$,
$\hat{y}_{\mathrm{cvr}}(i)$;\; $p_i$ &
  Predicted click probability, predicted conversion probability,
  and item display price &
  Reward-model heads and prediction-derived GMV proxy \\

$z_{\mathrm{CTR}},\;z_{\mathrm{CTCVR}},\;z_{\mathrm{GMV}}$ &
  Prediction-derived list-level business-score components;
  $z_{\mathrm{GMV}}$ uses item display prices rather than
  realized transaction amounts &
  Reward model \\

$P_k$;\; $q_k=P_k(z_k)$ \newline
$0\leq q_k\leq1$ &
  Reward-model-specific empirical percentile map and normalized
  component &
  Reward model; $P_k$ distinct from $P_{\mathrm{amt}}$;
  $q_k$ distinct from retrieval quota $q_i$ \\

$S_{\mathrm{RM}}(a)$ \newline
$=w_1^{\mathrm{RM}}\,q_{\mathrm{CTR}}$ \newline
${}+w_2^{\mathrm{RM}}\,q_{\mathrm{CTCVR}}$ \newline
${}+w_3^{\mathrm{RM}}\,q_{\mathrm{GMV}}$ &
  Learned-reward score combining percentile-normalized
  prediction-derived components with reward-model-specific
  mixing coefficients &
  GRPO reward \\

\addlinespace
%% ---- SFT and Wilson selection ----
\multicolumn{3}{l}{\emph{SFT and Wilson selection}}\\
\addlinespace

$w_{u,a},\;t_{u,a},\;l_{u,a}$ &
  Pairwise win, tie, loss counts &
  SFT substage 2; $w_{u,a}$ distinct from fusion weight $w$ \\

$n_{u,a}$;\; $\hat{p}_{u,a}$ &
  Total comparisons; empirical adjusted win rate &
  SFT substage 2 \\

$\mathrm{LB}(u,a)$;\; $z$;\; $\alpha$ &
  Wilson lower bound; standard-normal quantile
  ($z\!=\!1.96$); confidence parameter &
  SFT substage 2 \\

$\mathcal{A}_u$;\; $a_u^+$ &
  Evaluated configuration set for user $u$; selected SFT target &
  SFT substage 2 \\

\addlinespace
%% ---- FAPO ----
\multicolumn{3}{l}{\emph{FAPO (Format Anchored Preference Optimization)}}\\
\addlinespace

$S_{\mathrm{pay}}(y\mid x)$ &
  Purchase-only business score retaining only purchase-related
  NDCG and hit-rate contributions; Eq.~\eqref{eq:purchase_only} &
  FAPO pair construction \\

$\lambda_N$;\; $\lambda_H$ &
  Non-negative weighting coefficients for the NDCG and hit-rate
  components of $S_{\mathrm{pay}}$; scenario-specific,
  held fixed during training &
  FAPO pair construction; distinct from DNN loss weights
  $\lambda_1,\lambda_2$ \\

$a_g$;\; $\mathcal{G}$ &
  Per-group hit-rate weight and matching-group set
  (item, category, same-product) &
  FAPO pair construction; $a_g$ distinct from action~$a$ \\
$\mathcal{D}_{\mathrm{pref}}=\{(x,y^+,y^-)\}$ &
  Preference dataset with chosen / rejected responses &
  FAPO training \\

$p_s(y)$;\; $\bar{p}_s$ &
  Historical posterior V2P of orchestration in scenario $s$;
  scenario mean &
  Posterior-guided weighting \\

$\kappa_s$ &
  Scenario-specific separation threshold &
  Posterior-guided weighting \\

$\widehat{\Delta}_{\mathrm{post}}$;\;
$0.2\leq w_{\mathrm{pref}}\leq3.0$ &
  Normalized posterior gap; pair-level preference weight &
  FAPO loss; $w_{\mathrm{pref}}$ distinct from other uses of $w$ \\

$\mathcal{A}(y)$ &
  ActionMask: marked token positions for the current preference pair &
  FAPO; distinct from action space $\mathcal{A}$ \\

$g_\theta^{\mathcal{A}}(y\mid x)$;\;
$o_\theta^{\mathcal{A}}(y\mid x)$ &
  Action-restricted mean log-probability; its odds &
  FAPO ActionMask \\

$\mathcal{L}^{\mathrm{gen}}$;\;
$\mathcal{L}_{\mathrm{FAPO}}$;\; $\beta$ &
  Generation loss; FAPO objective;
  preference--generation balance coefficient &
  FAPO training ($\beta\!=\!0.5$) \\

$\sigma$ &
  Sigmoid function &
  FAPO loss; distinct from snapshot $\sigma_x$ \\

\addlinespace
%% ---- GRPO and Sandbox ----
\multicolumn{3}{l}{\emph{GRPO and Sandbox}}\\
\addlinespace

$\sigma_x$ &
  Request snapshot (frozen meta-parameters and behavior sequence) &
  Sandbox replay; distinct from sigmoid $\sigma$ \\

$\hat{F}(x,a;\sigma_x)$ &
  Sandbox pipeline replay under snapshot &
  GRPO exploration \\

$\pi_{\theta_{\mathrm{old}}}$;\; $\pi_{\mathrm{ref}}$ &
  Rollout policy; reference policy (both initialized from FAPO) &
  GRPO \\

$x_i$;\; $y^{(i,j)}$;\; $G\!=\!8$;\; $T\!=\!1.5$ &
  $i$-th request; $j$-th response in group;
  selected group size; sampling temperature &
  GRPO \\

$\hat{A}^{(i,j)}$ &
  Group-relative advantage &
  GRPO \\

$\rho_t^{(i,j)}$;\; $\bar{\rho}_t^{(i,j)}$ &
  Token-level importance ratio; clipped ratio &
  GRPO \\

$\varepsilon_{\mathrm{low}}\!=\!0.2$;\;
$\varepsilon_{\mathrm{high}}\!=\!0.28$ &
  Asymmetric clipping bounds &
  GRPO \\

$\beta_{\mathrm{KL}}\!=\!0.01$;\; $D_{\mathrm{KL}}$ &
  KL penalty coefficient; KL divergence &
  GRPO \\

$\mathbb{E}_{\mathrm{rollout}}$;\; $\mathcal{J}(\theta)$ &
  Expectation over rollout groups; GRPO training objective &
  GRPO; $\mathcal{J}$ distinct from business objective $J$ \\

\addlinespace
%% ---- DNN baseline ----
\multicolumn{3}{l}{\emph{DNN baseline}}\\
\addlinespace

$Q(x,a)=f_\theta\!\bigl(\phi(x),\psi(a)\bigr)$ &
  Pre-sigmoid orchestration logit;
  $f_\theta$: DNN scorer &
  DNN route; $\theta$ here denotes DNN parameters,
  not the LLM policy \\

$\phi(x)$;\; $\psi(a)$ &
  User representation; orchestration representation &
  DNN route; $\psi(a)$ distinct from reward-model parameter $\psi$ \\

$r$;\; $0\leq r\leq1$ &
  Direct bounded composite business-score target;
  $r=S_{\mathrm{comp}}(a)$ &
  DNN pointwise training; used with no additional
  normalization or clipping \\

$a_{\mathrm{win}},\;a_{\mathrm{lose}}$ &
  Superior / inferior orchestrations within a request &
  DNN pairwise loss \\

$\mathcal{L}_{\mathrm{reg}}$;\;
$\mathcal{L}_{\mathrm{pair}}$;\;
$\mathcal{L}
=\lambda_1\mathcal{L}_{\mathrm{pair}}
+\lambda_2\mathcal{L}_{\mathrm{reg}}$ &
  Pointwise BCE fitting loss on
  $\sigma\!\bigl(Q(x,a)\bigr)$ against the direct bounded
  composite-score target $r$; pairwise ranking loss on
  pre-sigmoid logit differences; combined objective
  ($\lambda_1\!=\!\lambda_2\!=\!1$) &
  DNN training \\

\addlinespace
%% ---- Evaluation metrics ----
\multicolumn{3}{l}{\emph{Evaluation metrics}}\\
\addlinespace

$\mathrm{V2P}$;\; $\mathrm{uCTR}$;\;
$\mathrm{pCTR}$;\; $\mathrm{GMV}$ &
  Visit-to-purchase rate; user CTR; page CTR;
  gross merchandise value &
  Online A/B evaluation; $\mathrm{GMV}$ here denotes the
  observed monetary business metric, not a percentile reward \\

$\mathrm{CTR}$;\; $\mathrm{CVR}$;\; $\mathrm{CTCVR}$ &
  Click-through rate; conversion rate;
  click-through conversion rate &
  Model-prediction heads and online metrics \\

Top-1 Hit Rate;\; Win-or-Tie Rate &
  Whether DNN picks the best candidate;
  fraction matching or exceeding LLM &
  DNN offline evaluation \\

\end{longtable}
}

\section{Action Space Configuration}
\label{sec:action_space_configuration}
This appendix provides representative configuration entries
referenced in Section~\ref{sec:action_space}.
The seven formal dimensions defined in Table~\ref{tab:action} are
$\mathcal{A} = \mathcal{A}_{\text{pipeline}} \times
              \mathcal{A}_{\text{retrieval}} \times
              \mathcal{A}_{\text{quota}} \times
              \mathcal{A}_{\text{ctr}} \times
              \mathcal{A}_{\text{cvr}} \times
              \mathcal{A}_{\text{fusion}} \times
              \mathcal{A}_{\text{rerank}}$.
Each subsection below details one or more of these dimensions,
followed by supplementary pre-ranking and a redacted
structured-action example.

%% ---------------------------------------------------------------
\subsection{Pipeline Modes}
\label{app:pipeline-modes}

As summarized in Table~\ref{tab:pipeline-modes}, MARS exposes three
pipeline topologies with different computational depths by skipping
selected stages. The agent selects exactly one mode for each request.

\begingroup
\floatplacement{table}{H}

\begin{table}
\caption{Pipeline mode configurations.}
\label{tab:pipeline-modes}

\begin{tabular*}{\textwidth}
{@{\extracolsep{\fill}}ll@{}}
\toprule
Mode & Stages \\
\midrule
Full
& retrieval $\to$ pre-ranking $\to$ ranking $\to$ re-ranking \\
Streamline
& retrieval $\to$ ranking $\to$ re-ranking \\
Minimal
& retrieval $\to$ ranking \\
\bottomrule
\end{tabular*}
\end{table}

\endgroup

In each scenario, the observed mode distribution is approximately \textsc{Full} $79\%$, \textsc{Streamline} $1\%$, and \textsc{Minimal} $20\%$.
When a stage is skipped, its selection and weight fields remain
present and are set to empty strings (\texttt{""}).
For \texttt{streamlinedpipeline}, \texttt{prerank} and
\texttt{prerank\_weights} are empty, while re-ranking remains active.
For \texttt{minimalpipeline}, \texttt{prerank},
\texttt{prerank\_weights}, \texttt{rerank}, and
\texttt{rerank\_weights} are empty.
For \texttt{fullpipeline}, both stages are active.

%% ---------------------------------------------------------------
\clearpage
\subsection{Retrieval Channels}
\label{app:retrieval-channels}

The full retrieval inventory comprises 20~catalogue entries organized
into three scenario groups; Table~\ref{tab:retrieval-inventory} shows
four representative entries.
At action time the agent selects a \emph{scenario-valid subset} of
channels from the pool available for the current business context,
so the per-request active set is a strict subset of the catalogue.
Catalogue keys are opaque identifiers assigned for compatibility with
existing configurations and are not required to be consecutive.
The \emph{Scenario} column indicates the business context in which each
channel is available, while the \texttt{recall\_num} and
\texttt{trigger\_num} columns indicate whether the corresponding
parameter is exposed to the agent for that channel.

\begin{table}[ht]
\centering
\caption{Representative retrieval-channel entries and exposed parameters.
         The full inventory comprises 20~entries; 16~additional entries
         covering further General, ShangouPostPay, and Dacu channels
         are omitted for brevity.}
\label{tab:retrieval-inventory}
\small
\begin{tabular}{lllcc}
\toprule
Key & Name & Scenario & \texttt{recall\_num} & \texttt{trigger\_num} \\
\midrule
\multicolumn{5}{l}{\emph{General (2 of 9 entries shown)}} \\
\texttt{recall\_1}  & i2i\_recall (item-to-item CF)        & General & \checkmark & \checkmark \\
\texttt{recall\_4}  & mind\_recall (multi-interest vectors) & General & \checkmark &            \\
\midrule
\multicolumn{5}{l}{\emph{ShangouPostPay (1 entry)}} \\
\texttt{recall\_10} & ShangouPostPay sid2i                  & ShangouPostPay & \checkmark & \checkmark \\
\midrule
\multicolumn{5}{l}{\emph{Dacu (1 of 10 entries shown)}} \\
\texttt{recall\_11} & WYSIWYG recall                        & Dacu & \checkmark &            \\
\bottomrule
\end{tabular}
\end{table}

%% ---------------------------------------------------------------
\subsection{Per-Channel Quotas}
\label{app:quotas}

For every selected retrieval channel the agent specifies
a \texttt{recall\_num} parameter that controls the number of
candidates returned.
Quotas are expressed as perturbations of a system default:

\begin{equation}
  q_c = q_c^{\text{default}} \times (1 + \delta_c),
  \qquad \delta_c \in [-0.2,\; +0.2].
  \label{eq:quota-perturbation}
\end{equation}

Table~\ref{tab:quota-defaults} lists known default quotas for one of the scenarios; defaults for other scenarios follow a
similar magnitude but are injected via per-scenario expert knowledge
at prompt time.

\begingroup
\floatplacement{table}{H}
\begin{table}
\centering
\caption{Default retrieval quotas for one of the scenarios.
         The agent may perturb each quota by at most $\pm 20\%$
         of the default value.}
\label{tab:quota-defaults}
\begin{tabular}{lccc}
\toprule
Channel & Default $q_c^{\text{default}}$ & Min ($-20\%$) & Max ($+20\%$) \\
\midrule
\texttt{i2i\_recall}    & 800 & 640 & 960 \\
\texttt{divi2i\_recall}  & 600 & 480 & 720 \\
\texttt{mnr\_recall}     & 500 & 400 & 600 \\
\texttt{Rhot\_recall}    & 300 & 240 & 360 \\
\texttt{mind\_recall}    & 300 & 240 & 360 \\
\texttt{…}    & … & … & … \\
\bottomrule
\end{tabular}
\end{table}
\endgroup

%% ---------------------------------------------------------------
\subsection{CTR and CVR Models}
\label{app:ctr-cvr-models}

The agent selects one CTR model and one CVR model for the ranking
stage, overriding the system defaults. The following tables show
representative catalogue entries using anonymized publication labels;
they do not enumerate the accepted wire-format values of
\texttt{rank\_ctr} and \texttt{rank\_cvr}.

\begin{table}[h]
\centering
\caption{Representative CTR model catalogue entries
using publication model labels (non-deployable placeholders).}
\label{tab:ctr-models}
\begin{tabular}{lll}
\toprule
Catalogue key & Publication label & Scenario \\
\midrule
\texttt{rank\_1} & Model-D\textsubscript{CTR} & Baiyibutie  \\
\texttt{rank\_2} & Model-C\textsubscript{CTR} & Baiyibutie  \\
…              & … & … \\
\bottomrule
\end{tabular}
\end{table}

\begin{table}[h]
\centering
\caption{Representative CVR model catalogue entries
using publication model labels (non-deployable placeholders).}
\label{tab:cvr-models}
\begin{tabular}{lll}
\toprule
Key & Model & Scenario \\
\midrule
\texttt{rank\_3} & Model-C\textsubscript{CVR} & Baiyibutie \\
\texttt{rank\_4} & Model-B\textsubscript{CVR} & Baiyibutie \\
…              & … & … \\
\bottomrule
\end{tabular}
\end{table}

%% ---------------------------------------------------------------
\subsection{Fusion Weights}
\label{app:fusion-weights}

At each scoring stage the final ranking score is computed as a
weighted combination of predicted click-through rate, conversion rate,
and item price:
\begin{equation}
  \text{score}(i)
  = \hat{y}_{\text{ctr}}(i)^{w_{\text{ctr}}}
    \cdot \hat{y}_{\text{cvr}}(i)^{w_{\text{cvr}}}
    \cdot p_i^{w_{\text{price}}}.
  \label{eq:fusion}
\end{equation}

The agent specifies each weight as a discrete perturbation of the
stage-specific default:
\begin{equation}
  w = w^{\text{default}} + k \times 0.05,
  \qquad k \in \{-6, -5, \ldots, 0, \ldots, 5, 6\},
  \label{eq:weight-discretization}
\end{equation}
defining 13~nominal offsets per weight parameter. The valid candidate
set retains only values satisfying non-negativity and scenario-specific
constraints.

\begin{table}[h]
\centering
\caption{Fusion weight defaults for one of the scenarios and ranges by pipeline stage.
         Each weight is perturbed independently via
         Eq.~\eqref{eq:weight-discretization}.}
\label{tab:fusion-weights}
\begin{tabular}{lccc}
\toprule
Stage & $w_{\text{ctr}}^{\text{default}}$
      & $w_{\text{cvr}}^{\text{default}}$
      & $w_{\text{price}}^{\text{default}}$ \\
\midrule
Pre-ranking  & 1.50 & 0.75 & 0.00 \\
Ranking      & 1.13 & 1.00 & 0.10 \\
\bottomrule
\end{tabular}
\end{table}

\noindent For example, the ranking-stage
$w_{\text{ctr}}$ ranges from $1.13 - 6 \times 0.05 = 0.83$ to
$1.13 + 6 \times 0.05 = 1.43$ in steps of $0.05$.

%% ---------------------------------------------------------------
\subsection{Re-Ranking}
\label{app:reranker}

Table~\ref{tab:reranker} shows representative re-ranking catalogue
entries rather than an exhaustive list of values accepted by the
serving system. For requests that include re-ranking, the agent
selects one scenario-valid strategy.

\begin{table}
\centering
\caption{Representative re-ranking catalogue entries.}
\label{tab:reranker}
\begin{tabular}{lll}
\toprule
Catalogue key & Strategy & Scenario \\
\midrule
\texttt{rerank\_1} & DPP & General \\
\texttt{rerank\_4} & Next-gen re-ranking & Baiyibutie \\
\multicolumn{3}{c}{\ldots} \\
\bottomrule
\end{tabular}
\end{table}

\noindent\textbf{Next-gen weights.}
\texttt{rerank\_4} exposes four dedicated fusion weights with the
following defaults:
\begin{itemize}
  \item $w_{\mathrm{ctr}}=0$,
        $w_{\mathrm{cvr}}=0$,
        $w_{\mathrm{GMV}}=0$, and
        $w_{\mathrm{rank}}=1$.
\end{itemize}

\noindent
By default, only the upstream ranking-model score contributes to the
Next-gen score. These weights follow the same discrete perturbation
rule in Eq.~\eqref{eq:weight-discretization} as the fusion weights in
Section~\ref{app:fusion-weights}.

%% ---------------------------------------------------------------
\subsection{Pre-Ranking}
\label{app:preranking}

Pre-ranking is executed only under the \emph{Full} pipeline mode.
Two pre-ranking models are shown as follows:

\begin{table}[h]
\centering
\caption{Pre-ranking models and exposed parameters.}
\label{tab:preranking}
\begin{tabular}{lllccc}
\toprule
Key & Model & Scenario
    & \texttt{ctr\_w} & \texttt{cvr\_w} & \texttt{price\_w} \\
\midrule
\texttt{prerank\_1} & Two-tower pre-ranking
                    & Baiyibutie
                    & \checkmark & \checkmark & \checkmark \\
\texttt{prerank\_2} & Residual-quantized pre-ranking
                    & Dacu
                    & \checkmark & \checkmark & \checkmark \\
\bottomrule
\end{tabular}
\end{table}

\noindent Pre-ranking fusion follows the same multiplicative score-fusion formula as
ranking (Eq.~\eqref{eq:fusion}), with defaults
$w_{\text{ctr}} = 1.5$, $w_{\text{cvr}} = 0.75$,
$w_{\text{price}} = 0$
(see Table~\ref{tab:fusion-weights}).

%% ---------------------------------------------------------------
\subsection{Structured Action Example}
\label{app:action-example}

Listing~\ref{lst:action-json} shows a redacted structured-action
example for a subsidy-scenario request under the \emph{Full}
pipeline mode. It illustrates the serving wire-format field
structure rather than an action that can be deployed unchanged.
The model and scene identifiers are publication placeholders;
deployment requires identifiers accepted by the serving system.

The system prompt presents pipeline modes using human-readable
spaced names (\texttt{full pipeline}, \texttt{streamlined pipeline},
\texttt{minimal pipeline}), whereas the model output uses
concatenated identifiers (\texttt{fullpipeline},
\texttt{streamlinedpipeline}, \texttt{minimalpipeline}).
This mapping is established through SFT training data;
the consumption-side validator accepts only the concatenated form.

\begingroup

\lstset{
  language={},
  literate={},
  basicstyle=\small\ttfamily\color{black},
  keywordstyle=\color{black},
  stringstyle=\color{black},
  commentstyle=\color{black},
  identifierstyle=\color{black},
  emphstyle=\color{black},
  numberstyle=\tiny\color{black},
  rulecolor=\color{black},
  backgroundcolor=\color{white},
  breaklines=true,
  columns=fullflexible,
  showstringspaces=false,
  frame=single,
  captionpos=t,
  escapeinside={(*@}{@*)},
  caption={Publication-format structured-action excerpt. Scene and model identifiers are placeholders, and the request-specific explanation is omitted.},
  label={lst:action-json}
}

\begin{lstlisting}
{
  "global_thought": "[REQUEST_SPECIFIC_EXPLANATION_OMITTED]",
  "scene": "(*@\char91@*)SCENE_ID(*@\char93@*)",
  "intent": "bundle-building",
  "tools": "fullpipeline",
  "actions": {
    "recall": (*@\char91@*)
      "i2i_recall",
      "divi2i_recall",
      "mind_recall",
      "Rhot_recall",
      "mnr_recall"
    (*@\char93@*),
    "recall_weights": (*@\char91@*)
      "880",
      "660",
      "330",
      "330",
      "550"
    (*@\char93@*),
    "prerank": "TwoTower",
    "prerank_weights": "1.5#0.75#0",
    "rank_ctr": "Model-A_CTR",
    "rank_cvr": "Model-A_CVR",
    "rank_weights": "1.13#1.15#0.1",
    "rerank": "Next-gen",
    "rerank_weights": "0#0#0.1#1"
  }
}
\end{lstlisting}

\endgroup

\noindent\textbf{Validation.}
Validation is applied to the unredacted action rather than to the
publication placeholders shown above. The required checks include:
\begin{itemize}
  \item JSON schema conformance and presence of all required fields;

  \item exact pipeline identifiers:
        \texttt{fullpipeline},
        \texttt{streamlinedpipeline}, or
        \texttt{minimalpipeline};

  \item scenario-specific allow-list membership for retrieval
        channels, pre-ranking models, CTR and CVR models, and
        re-ranking strategies;

  \item positional alignment and scenario-specific bounds for
        retrieval channels and quotas;

  \item valid component counts, discrete candidate values,
        non-negativity, and scenario-specific constraints for
        active-stage weights;

  \item stage compatibility: skipped-stage fields remain present
        and contain empty strings.
        Under \texttt{streamlinedpipeline},
        \texttt{prerank} and \texttt{prerank\_weights} are empty.
        Under \texttt{minimalpipeline},
        \texttt{prerank}, \texttt{prerank\_weights},
        \texttt{rerank}, and \texttt{rerank\_weights} are empty.
\end{itemize}

Actions that fail validation are rejected and are not written to the
cache. If a valid cached action exists, it remains available for
subsequent requests; otherwise, the fallback rules in
Section~\ref{sec:orchestration_workflow} apply.

\section{Prompt Details}
\label{app:prompt_details}
% ============================================================
% Appendix D — Prompt Details (Privacy-Redacted Version)
% ============================================================
% Prerequisites: ensure the following packages are loaded in
% the document preamble:
%   \usepackage{listings}
%   \usepackage{xcolor}
% ============================================================

\lstdefinestyle{promptstyle}{
  basicstyle=\scriptsize\ttfamily,
  breaklines=true,
  columns=fullflexible,
  showstringspaces=false,
  frame=single,
  captionpos=b,
}

%% ----- D.1 Overview -----
\subsection{Overview}
\label{app:prompt_overview}
The system prompt in Section~\ref{app:baiyibutie_prompt} is a publication excerpt derived from the production template: internal identifiers are replaced by publication placeholders, while parameter defaults and action constraints reproduce the configuration values documented in Appendix~\ref{sec:action_space_configuration}. The user prompt in Section~\ref{app:baiyibutie_user_prompt} is a fully synthetic request example. Its profile values, timestamps, behavioral records, search queries, and memory statistics are synthetically constructed and do not correspond to a production user. Omitted content is explicitly
marked.

The system message defines the agent role and objective, describes the input
schema, provides the knowledge base and action boundaries, and constrains the
output.  The following user message supplies the concrete scene context, user
profile and model features, behavior sequence, search context, and long-term
memory for one request.  Section~\ref{app:prompt_assembly} explains how the two
messages are assembled, and Section~\ref{app:prompt_template} summarizes their
structural components.

%% ----- D.2 Baiyibutie System Prompt -----
\subsection{Baiyibutie System Prompt}
\label{app:baiyibutie_prompt}

Listing~\ref{lst:baiyibutie_system_prompt} presents an
English-translated, publication excerpt of the Baiyibutie system prompt.
Spaced pipeline names in this excerpt are prompt-side display names,
not serving-side \texttt{tools} identifiers. Their correspondence to
machine identifiers is established through SFT, as described in
Appendix~\ref{app:action-example}; no serving-side alias normalization is implied.

Model labels in this excerpt are publication model labels (non-deployable placeholders). The output-format block is a schematic
template that illustrates field structure, not an executable action.
Ellipses indicate omitted entries and do not define serving allow-lists.

\begin{lstlisting}[style=promptstyle,
  caption={Baiyibutie scene system prompt
(English-translated, publication excerpt; section structure and core
execution constraints preserved).},
  label={lst:baiyibutie_system_prompt}]
You are a senior recommendation system algorithm expert, specializing in strategy optimization for Alibaba's MARS (Multi-stage Agentic Recommendation System).
## Objective
Your task is to produce a high-quality system orchestration strategy based on the given [User Profile], [Behavior Sequence], and [Scene Context].

---

## Input Data Schema (User Side)

### 0. Scene Context
- Scene type: the shopping scene the user is currently in
- Scene description: Subsidy-driven scene with price-sensitive users and conversion-focused recommendation objectives. ... (Detailed scene-specific guidance omitted.)
- Scene ID: [SCENE_ID]
- User's current page
- Current time
- Season
- Intent

### 1. User Profile Features (User Profile)
- User ID
- Membership level
- Spending power level
... (4 additional profile-field definitions omitted.)

#### Model Feature Importance Reference
Use the selected high-importance features supplied in the user message.

### 2. Cross-Scene Behavior Sequence (Behavior Sequence)
- Format: Markdown table containing `[time, timestamp, dwell time, action (click/view/skip/add_cart/buy), item title, category name, shop name]`.

### 3. User Search Context (Search Context)
- Format: Markdown table or list containing `[time (YYYY-MM-DD HH:MM), search query]`
- Note: Records the user's search behavior on the main search page, reflecting active intent and interest preferences.

### 4. Memory Context (Memory Context)
- Long-term memory: annual purchase activity and recent cart-add/purchase/view activity.
... (Additional memory-field definitions omitted.)

---

## Knowledge Base and Constraints (Assistant Side)

The assistant's response must strictly follow the MARS prior knowledge below:
**Note**: Cost represents the tool invocation cost. When selecting tools, the assistant must understand the user profile and recent behavior, then make decisions by jointly considering relevance, performance, and invocation cost. Cost is a value in (1-10); higher means more expensive.

### 1. Pipeline Tools
The recommendation system offers three pipeline modes; the `tools` field must be one of the following:

| Pipeline Mode | tools Value | Stages |
|----------|----------|----------|
| Full pipeline | `full pipeline` | retrieval -> pre-ranking -> ranking -> re-ranking |
| Streamlined pipeline | `streamlined pipeline` | retrieval -> ranking -> re-ranking |
| Minimal pipeline | `minimal pipeline` | retrieval -> ranking |

### 2. Action Space

##### Metric Definitions
| Metric | Meaning |
|------|------|
| **pvr** | Exposure share of this retrieval channel |
| **ctr** | Click-through rate |
| **cvr** | Conversion rate |
| **price** | Item price |
| **GMV weight** | Weight of ctr x cvr x price |

##### Recall (Retrieval)
| Action Name | Description | Key Parameters | Key Metric | Cost |
|-----------|-------------|----------------|----------------|------|
| `i2i_recall` | Item-to-Item collaborative filtering; retrieves similar items based on user historical behavior. | `w1`: quota | ctr, cvr | 7 |
| `mind_recall` | Multi-interest vector retrieval. | `w1`: quota | ctr, cvr | 7 |
... (5 additional recall-channel entries omitted.)

**Retrieval tool selection rules**: Select 3-7 scenario-valid retrieval channels.

**Baseline retrieval quota (Baiyibutie scene)**:
| Retrieval Channel | Default quota |
|----------|----------|
| `i2i_recall` | 800 |
| `mind_recall` | 300 |
... (3 additional baseline-quota entries omitted.)
Quota must be a positive integer; perturbation range must not exceed 20% of the baseline value.

##### Pre-Rank (Pre-Ranking)
| Action Name | Description | Key Parameters | Cost |
|-----------|-------------|----------------|----------------|
| `TwoTower` | Two-tower pre-ranking model; filters candidate sets balancing efficiency and accuracy. | `w1`, `w2`, `w3` | 7 |
Overall output weight: `w1`: ctrWeight, `w2`: cvrWeight, `w3`: priceWeight. Defaults: 1.5, 0.75, 0

##### Rank (Ranking)
| Action Name | Description | Key Parameters | Cost |
|-----------|-------------|----------------|-------------|
| `Model-A_CTR` | CTR model; price-consistent CTR with pricing engine optimization. | `w1`, `w2`, `w3` | 7 |
| `Model-E_CTR` | CTR model; lightweight CPU-only model. | `w1`, `w2`, `w3` | 3 |
... (3 additional CTR model entries omitted.)
| `Model-A_CVR` | CVR model; price-consistent CVR with pricing engine optimization. | `w1`, `w2`, `w3` | 7 |
... (3 additional CVR model entries omitted.)
Overall output weight: `w1`: ctrWeight, `w2`: cvrWeight, `w3`: priceWeight. Defaults: 1.13, 1, 0.1
Select one CTR model and one CVR model for output.

##### Re-Rank (Re-Ranking)
| Action Name | Description | Key Parameters | Cost |
|-----------|-------------|----------------|------|
| `DPP` | DPP-based re-ranking. | `w1`, `w2`, `w3` | 6 |
| `Next-gen` | Generation-based re-ranking using CTR, CVR, GMV, and ranking-score weights. | `w1`, `w2`, `w3`, `w4` | 8 |
...

`Next-gen` uses `w1` for CTR, `w2` for CVR, `w3` for GMV, and `w4` for the upstream ranking score. The default values are 0, 0, 0, and 1.
DPP uses `w1` for CTR, `w2` for CVR, and `w3` for GMV.

---

## Sample Generation Principles
1. The number of retrieval channels should be flexibly selected based on scene complexity, typically 3-7 channels.
2. Weights must be strings in hash-separated format: pre-ranking and ranking use CTR#CVR#Price; DPP re-ranking uses CTR#CVR#GMV; Next-gen re-ranking uses CTR#CVR#GMV#Rank.
3. **Parameter perturbation and fluctuation rules**:
   - Retrieval channel quota: must be a positive integer, perturbed from the baseline default; the perturbation range must not exceed 20% of the baseline value.
   - Other weight parameters (pre-ranking/ranking/re-ranking): perturb in steps of 0.05 from the default (new value = default + k x 0.05), where k is an integer in [-6, 6]; weights are not required to sum to 1.
5. Separation of explanation and execution:
   - `global_thought` contains the explanation
   - `actions` contains the action parameters
6. tools must be exactly one of: minimal pipeline | streamlined pipeline | full pipeline
7. You must wrap your output in ```json``` and ensure it is valid JSON.
8. **Retrieval channel selection constraint**: The selected retrieval channels must be chosen from the
scenario-specific allowed channel set and should cover the major
baseline channels. Unless there is strong confidence, major baseline
channels should not be removed.
9. Weight adjustment guidelines:
* High cart-add but low purchase-conversion users: increase cvrWeight to promote conversion; adjust priceWeight based on coupon usage; focus retrieval on highly relevant items; maintain content appeal.
... (3 additional user-segment guidelines omitted.)
* Weights must not be negative.

## Output Format

```json
{"global_thought":"your reasoning","scene":"scene id","intent":"intent type from user input","tools":"minimal pipeline or streamlined pipeline or full pipeline","actions":{"recall":["recall_channel_1","recall_channel_2"],"recall_weights":["quota_1","quota_2"],"prerank":"pre-ranking model name", "prerank_weights": "xx#xx#xx","rank_ctr":"CTR model name","rank_cvr":"CVR model name","rank_weights":"xx#xx#xx", "rerank": "re-ranking model name", "rerank_weights": "xx#xx#xx or xx#xx#xx#xx (Next-gen)"}}
```
**Note**:
Retain all stage fields in the JSON.
- For the streamlined pipeline, set `prerank` and `prerank_weights` to empty strings; re-ranking remains active.
- For the minimal pipeline, set `prerank`, `prerank_weights`, `rerank`, and `rerank_weights` to empty strings.
- For the full pipeline, both pre-ranking and re-ranking are active.
### `global_thought`
Contains **explanatory text**, including but not limited to:
- User cross-scene behavior analysis
- Rationale for strategy selection
- Constraints and trade-off explanations
- Keep the output concise, 3-5 sentences maximum, without Markdown formatting.

### `actions`
Contains only **executable parameters**; keep them clear, stable, and parseable.
\end{lstlisting}

%% ----- D.3 Baiyibutie User Prompt — Request Example -----
\subsection{Baiyibutie User Prompt --- Fully Synthetic Request Example}
\label{app:baiyibutie_user_prompt}

Listing~\ref{lst:baiyibutie_user_prompt} presents a fully synthetic
request example illustrating the user-message structure. All profile
values, timestamps, behavioral records, search queries, and memory
statistics are synthetically constructed and do not correspond to any
production user or real trajectory. They demonstrate the field layout
and representative value ranges only. Omitted content is explicitly
marked.

\begin{lstlisting}[style=promptstyle,
  caption={Fully synthetic Baiyibutie user-prompt example, translated into English. This is a selective publication excerpt; omission of a field from the example does not imply its exclusion from the runtime input.},
  label={lst:baiyibutie_user_prompt}]
## Scene Context
Current scene is Baiyibutie

- Scene type: Baiyibutie
- Scene description: 
- Scene ID: [SCENE_ID]
- Page: 0
- Time: Thursday 14:22
- Intent: casual-browsing intent
- Season: Spring

## User Profile
- User ID: [SYNTHETIC_USER]
- Spending power level: 3
- Membership: V1
... (additional profile and model-feature fields omitted; shown values are synthetic)

## Cross-Scene User Behavior Sequence
| No. | Time | Dwell Time(s) | Action | Item Title | Category | Shop Name |
|------|------|-------------|------|--------|--------|--------|
| 1 | 2d 5h ago | 6 | click | Wireless noise-canceling headphones | Headphones | [SHOP_P] |
... (8 more entries; all items are synthetic)
| 10 | 1d 3h ago | 3 | click | Spring running shoes (men) | Sports shoes | [SHOP_Q] |
... (7 more entries)
| 18 | 5h ago | 9 | add_cart | Portable Bluetooth speaker | Audio equipment | [SHOP_R] |

## User Search Context
| Time | Search Query |
|------|-----------|
| 2026-03-10 | Noise-canceling headphones |
... (3 more queries; all queries are synthetic)
| 2026-03-15 | Portable speaker Bluetooth |

## Memory Context
### Long-Term Memory
- 88VIP: 1
- Annual purchase count: ~120
- Annual purchase amount: ~8k
- Last 7 days cart-adds/purchases/views: 5/2/85
... (12 additional memory fields; values are synthetic)
\end{lstlisting}

%% ----- D.4 Message Structure and Assembly -----
\subsection{Message Structure and Assembly}
\label{app:prompt_assembly}

Each inference request is assembled as exactly two input messages in the
following order:
\begin{enumerate}
  \item \textbf{System message.} It establishes the recommendation-expert
        role, declares the expected input schema, provides the knowledge base,
        defines the available actions and parameter boundaries, and specifies
        the output constraints.
  \item \textbf{User message.} It supplies the current request instance,
        including its scene context, profile and model-feature values, behavior
        sequence, search history, and long-term memory.
\end{enumerate}
The schema description in the system message documents the fields expected in
the subsequent user message; it is not itself the user prompt.

%% ----- D.5 Prompt Template Structure -----
\clearpage\phantomsection
\subsection{Prompt Template Structure}
\label{app:prompt_template}

Table~\ref{tab:prompt_components} summarizes the components of both messages
and makes each component's message role explicit.
The table describes the runtime prompt; the system-prompt listing is a publication excerpt derived from the production template, and the user-prompt listing is a fully synthetic example.

\begin{table}[ht]
\centering
\caption{Structural components of the two-message MARS prompt.}
\label{tab:prompt_components}
\small
\begin{tabular}{@{}p{1.9cm}p{3.5cm}p{9.2cm}@{}}
\toprule
\textbf{Role} & \textbf{Component} & \textbf{Responsibility} \\
\midrule
System & Role and Objective
  & Assigns the recommendation-system expert role and states the task of producing an orchestration strategy from request context. \\
\addlinespace
System & Input Schema
  & Declares the expected categories and fields for scene context, user profile and model features, behavior sequence, search context, and memory context. This declaration describes the user input; it is not the request instance itself. \\
\addlinespace
System & Knowledge Base and Action Boundaries
  & Defines pipeline modes, available Recall, Pre-Rank, Rank, and Re-Rank capabilities, costs, metrics, model priorities, and baseline quotas and weights. \\
\addlinespace
System & Action-Generation Rules
  & Constrains channel selection, quota perturbations, weight formatting and perturbations, pipeline compatibility, and user-segment-dependent adjustments. \\
\addlinespace
System & Output Constraints
  & Requires a valid JSON result and defines the expected fields, separation of explanation from executable actions, and handling of stages omitted by the selected pipeline. \\
\addlinespace
User & Current Request Data
  & Instantiates the declared schema with the current scene, profile and model-feature values, behavior sequence, search records, and long-term memory. \\
\bottomrule
\end{tabular}
\end{table}

%% ----- D.6 Notes on Other Scenarios -----
\subsection{Notes on Other Scenarios}
\label{app:prompt_other_scenarios}

Other MARS scenarios use the same system-then-user message structure.  Based on
the available prompt materials, scenario-specific variation is limited to the
scene description, available capabilities, default configurations, and the
request data carried by the user message.  The message roles and assembly order
remain unchanged.

\end{document}